\documentclass[hidelinks,11pt,reqno]{article}

\usepackage{graphicx}
\usepackage{amssymb, amsthm, amsmath, dsfont, braket, mathtools, shadethm, xfrac}
\usepackage{cases}
\usepackage{amsfonts,exscale,latexsym,float}
\usepackage{authblk} 
\usepackage{geometry}
\usepackage{indentfirst,csquotes}
\usepackage[T1]{fontenc}
\usepackage[toc,page]{appendix}
\newtheorem{theorem}{Theorem}[]
\newtheorem{definition}{Definition}

\newtheorem{lemma}{Lemma}
\newtheorem{proposition}[theorem]{Proposition}

\theoremstyle{remark}
\newtheorem{remark}{Remark}

\makeatletter

\usepackage{subfiles}
\usepackage{paralist, hyperref, fancyhdr, etoolbox}
\usepackage{tikz}
\usetikzlibrary{positioning,calc,shapes,arrows}
\tikzset{
block/.style = {draw, fill=white, rectangle, minimum height=3em, minimum width=3em},
tmp/.style  = {coordinate}, 
sum/.style= {draw, fill=white, circle, node distance=1cm},
input/.style = {coordinate},
output/.style= {coordinate},
pinstyle/.style = {pin edge={to-,thin,black}
}
}

\usepackage[normalem]{ulem} 
\usepackage{amsmath,amsthm,amssymb,bm,bbm,dsfont,braket}
\usepackage[mathscr]{eucal}

\makeatother

\DeclareMathOperator{\Tr}{Tr}

\begin{document}

\title{ \bfseries
Maximal $\alpha$-Leakage for Quantum Privacy Mechanisms
}
\author[1]{Bo-Yu Yang \thanks{106yang@gmail.com}}
\author[1]{Hsuan Yu \thanks{yu.sherry3@gmail.com}}
\author[1--5]{Hao-Chung Cheng \thanks{haochung.ch@gmail.com}}

\affil[1]{\small \textit{Department of Electrical Engineering, National Taiwan University, Taipei 106, Taiwan (R.O.C.)}}
\affil[2]{\small \textit{Graduate Institute of Communication Engineering, National Taiwan University}}
\affil[3]{\textit{Department of Mathematics, National Taiwan University}}
\affil[4]{\textit{Hon Hai (Foxconn) Quantum Computing Center, New Taipei City 236, Taiwan (R.O.C.)}}
\affil[5]{\textit{Physics Division, National Center for Theoretical Sciences, Taipei 10617, Taiwan (R.O.C.)}}



\date{}
\maketitle

\begin{abstract}
    In this work, maximal $\alpha$-leakage is introduced to quantify how much a quantum adversary can learn about any sensitive information of data upon observing its disturbed version via a quantum privacy mechanism. 
    We first show that an adversary's maximal expected $\alpha$-gain using optimal measurement is characterized by measured conditional R\'enyi entropy.
	This can be viewed as a parametric generalization of K{\"o}nig \textit{et al.}'s famous guessing probability formula [\href{https://doi.org/10.1109/TIT.2009.2025545}{\textit{IEEE~Trans.~Inf.~Theory}, 55(9), 2009}].
    Then, we prove that the $\alpha$-leakage and maximal $\alpha$-leakage for a quantum privacy mechanism are determined by measured Arimoto information and measured R\'enyi capacity, respectively. 
    Various properties of maximal $\alpha$-leakage, such as data processing inequality and composition property are established as well.
    Moreover, we show that regularized $\alpha$-leakage and regularized maximal $\alpha$-leakage for identical and independent quantum privacy mechanisms coincide with $\alpha$-tilted sandwiched R\'enyi information and sandwiched R\'enyi capacity, respectively.
\end{abstract}

\section{Introduction} \label{sec:Introduction}
\subsection{Background}
In the era of Internet of Things, data are published and shared almost ubiquitously, which have significantly increased possible paths where private information can leak.
For instance, political preferences may be unveiled through movie ratings \cite{narayanan2008robust}.
Besides, publishing data while maintaining desirable privacy and utility guarantee has also come to issues in many fields such as business \cite{Hu2020cambridge}, healthcare \cite{Abouel2018medical}, and so on.
How can we prevent adversaries from inferring sensitive data via reverse engineering while preserving at least some levels of utility?
A common approach is to perturb the data by properly designing a \textit{privacy mechanism}: a stochastic map aiming to control the private information leakage to adversaries.
Conventionally, there are two mainstream approaches: (i) \textit{differential privacy} (DP) \cite{dwork2006differential, dwork2014algorithmic}, proposed by Dwork in 2006, guarantees the indistinguishability between neighboring data, addressing the paradox of learning less about an individual while gaining useful information about a population, which is further generalized to \textit{pufferfish privacy} \cite{kifer2014pufferfish} by Kifer \textit{et al.}~in 2014;
(ii) \textit{information-theoretic privacy}, which has been investigated in variant settings (e.g.~Shannon's mutual information \cite{shannon1949communication} and information bottleneck function \cite{tishby2000informationbottleneck}), adopts the viewpoints from statistics and information-theoretic security to study how the privacy is quantified \cite{du2012privacyStatInfer, sankar2013utility, asoodeh2016informationExtract, wang2017estimation}. 
Concurrently in 2017, Assodeh \textit{et al.}~proposed probability of correctly guessing \cite{asoodeh2017privacy, asoodeh2018estimation}, whereas Issa \textit{et al.}~proposed the \textit{maximal leakage} (MaxL) \cite{issa2019operational}, both aiming to depict how much sensitive information adversary can infer from its released version.
In 2018, Liao \textit{et al.}~further introduced $\alpha$-\textit{leakage} and \textit{maximal} $\alpha$-\textit{leakage} \cite{liao2019tunable} (both were extended to $(\alpha, \beta)$-leakage \cite{gilani2023alphabeta} later) to generalize the aforementioned quantities, so as to capture various adversarial actions with a tunable parameter $\alpha \in [0,\infty]$ \cite{liao2020WPSmaximalAlpha,sypherd2022alphaloss}. 
More recently, Stratonovich's value of information \cite{kamatsuka2022voi} and maximal $g$-leakage \cite{kurri2023operationalGenGain} are proposed to subsume maximal leakage and maximal $\alpha$-leakage, generalizing $\alpha$-gain/loss depictions to arbitrary gain functions \cite{KTH2023pointwise,KTH2023privateZero,KTH2023newPrivacyMechanism}.

When quantum devices become mature in the future, it is natural to consider protecting sensitive data via a \textit{quantum privacy mechanism}.
To this end, we may construct quantum privacy mechanisms via encoding sensitive classical data into quantum states. 
On the other hand, a quantum adversary may observe the encoded quantum states by applying certain quantum measurement.
Various methods and metrics have been proposed to help design quantum privacy mechanisms, including quantum differential privacy \cite{zhou2017differential, aaronson2019gentle, watkins2023qmlDP, angrisani2022DPamp, du2022quantumDPsparse} and quantum pufferfish privacy \cite{hirche2023quantum, nuradha2023pufferfish}; meanwhile, maximal quantum leakage \cite{farokhi2023maximal}, an extension of MaxL for quantum privacy mechanisms, has also been studied. 
Furthermore, Chen and Hanson \textit{et al.}~depict the guessing strategies of classical data encoded via quantum privacy mechanism to quantum states, which generalize the framework of guesswork proposed by Massey \cite{massey1994guessing} to quantum state discrimination problem in \cite{chen2014minimum, hanson2021guesswork}.


\subsection{Problem Formulation}

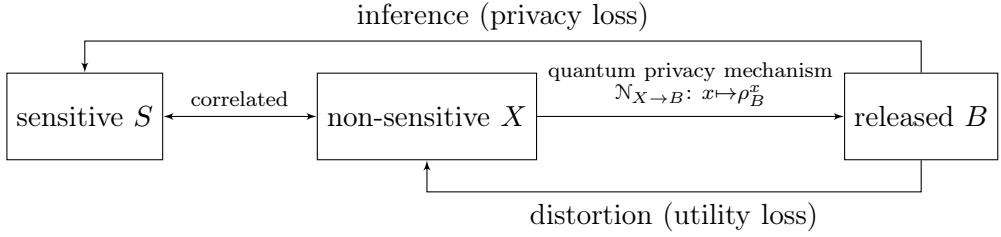
\begin{figure}[htbp]
  \centering
\begin{tikzpicture}[auto, node distance=4.5cm,>=latex']
    \node[block](S){sensitive $S$};
    \node[block, right of = S](X){non-sensitive $X$};
    \node[block, right of = X,  node distance=6.5cm](B){released $B$};
    \node[tmp, above of = S, node distance = 1cm](tmp1){};
    \node[tmp, below of = X, node distance = 1cm](tmp2){};
    \node[tmp, below of = B, node distance = 1cm](tmp3){};
    \node[tmp, above of = B, node distance = 1cm](tmp4){};
    \draw[<->] (S) -- node{ \scriptsize correlated}(X);
    \draw[->] (X) -- node[name=quantum]{$\substack{\textnormal{quantum privacy mechanism}\\ \mathscr{N}_{X\to B}:\ x \mapsto \rho_B^x}$
    }(B);
    \draw[<-](S) |- (tmp1) -- node{inference (privacy loss)}(tmp4) -| (B);
    \draw[->](B) |- (tmp3) -- node{distortion (utility loss)}(tmp2) -- (X);
    \node[tmp, below of = quantum, node distance = 3cm](tmp3){privacy mechanism};    
\end{tikzpicture}
\caption{Information leakage framework with quantum privacy mechanisms.
Here, systems $S$ and $X$ are classical, while system $B$ is quantum.
A quantum adversary may apply a quantum measurement on quantum system $B$ to infer data $X$ or $S$.
}
  \label{fig:IL_frame}
\end{figure}

In this work, we consider the problem of information leakage formulated as follows (Fig.~\ref{fig:IL_frame}).
Assume that random variable $X$ on a finite set $\mathcal{X}$ with probability distribution $p_X$ represents the non-sensitive data, which may be correlated with another random variable $S$, denoting some sensitive data.
A user wants to share non-sensitive $X$ with the service provider to gain utility while maintaining privacy, especially protecting information of sensitive $S$.
The user may apply a quantum privacy mechanism, namely, a classical-quantum channel $\mathscr{N}_{\mathcal{X} \to B}:x\mapsto \rho_B^x$, mapping each realization $x \in \mathcal{X}$ to a quantum state $\rho_B^x$ as a perturbation.\footnote{ If $\{\rho_B^x\}_x$ share a common eigenbasis, then the map $x\mapsto \rho_B^x$ reduces to classical privacy mechanism, i.e.~a stochastic transformation.}
In this manner, a fundamental question naturally arises:
\begin{equation*}
    \textit{How much information does a quantum system } B \textit{ leak about } S?
\end{equation*}

To observe the released quantum data $B$, adversaries can guess the non-sensitive data $X$ via a quantum measurement to obtain their inference $\hat{X}$ according to Born's rule \cite{gleason1957}:
\begin{equation}
    \Pr\{\hat{X} = X \mid X=x,B\} = \Tr[\rho_B^x \Pi_B^x],
\end{equation}
where $X-B-\hat{X}$ forms a classical-quantum-classical Markov chain.
Here, the collection $\{\Pi_B^x\}_{x\in\mathcal{X}}$ of positive semi-definite matrices  satisfying unity of resolution: $\sum_x \Pi_B^x = \mathds{1}_B$, i.e.~\textit{positive operator-valued measure} (POVM), represents a quantum measurement, which can be viewed as the quantum generalization of classical decisions.

Liao \textit{et al.}~characterized various classical adversarial actions by proposing $\alpha$-loss with $\alpha \in [1,\infty]$, interpolating between log-loss (for $\alpha = 1$) and 0-1 loss (also known as hard decision, for $\alpha = \infty$).
Recently, Sypherd \textit{et al.}~\cite{sypherd2022alphaloss} extended $\alpha$-loss to the range $(0,\infty]$ to include exponential loss (for $\alpha = \sfrac{1}{2}$) as well.
Since $\alpha$-leakage can be directly captured by the multiplicative gain increase upon an adversary observes the released data, motivated by Diaz \textit{et al.}~'s observation of maximal expected gain function \cite[Eq.~(8)]{diaz2019robustness}, we introduce the \textit{maximal expected $\alpha$-gain} of a quantum ensemble $\{p_X(x), \rho_B^x\}_{x\in\mathcal{X}}$
(Def.~\ref{def:max-alpha-gain}) as
\begin{equation}
    \mathsf{P}_\alpha(X|B)_\rho 
    :=  \sup_{\textsc{POVM}\, \{\Pi_B^x\}_x}\mathds{E}_{x\sim p_X}\left[\Tr\left[ \rho_B^x (\Pi_B^x)^\frac{\alpha-1}{\alpha} \right]\right], \quad \forall \alpha \in [1,\infty]
\end{equation}
to characterize how much an adversary correctly guesses the non-sensitive data $X$ when using an optimal measurement strategy for observing the released quantum system $B$.
If an adversary guesses $X$ without observing $B$ to obtain inference $\hat{X}$, we denote the corresponding (unconditional) maximal expected $\alpha$-gain (Def.~\ref{def:max-alpha-gain}) as
\begin{equation}
    \mathsf{P}_\alpha(X)_\rho := \sup_{p_{\hat{X}}:\hat{X} \perp X} \mathds{E}_{x\sim p_X}\left[ \Pr(\hat{X} = X | X = x)^{\frac{\alpha-1}{\alpha}} \right] = \sup_{p_{\hat{X}}} \mathds{E}_{x\sim p_X}\left[ {p_{\hat{X}}}(x)^{\frac{\alpha-1}{\alpha}} \right], \quad \forall \alpha \in [1,\infty],
\end{equation}
where the condition $\hat{X} \perp X$ of the maximizer in the first expression springs from conditional independence of Markov chain $X-B-\hat{X}$;
without the prior knowledge embedded in $B$, an adversary can only guess $\hat{X}$ independently.

In order to quantify how much information $B$ leaks about the non-sensitive $X$, we introduce the \textit{$\alpha$-leakage} (Def.~\ref{def:alpha-Leakage}) from non-sensitive data $X$ to quantum system $B$
as the multiplicative increase of maximal expected gain upon observing $B$, written as
\begin{equation}
    {\mathcal{L}}_{\alpha} (X \rightarrow B)_\rho := \frac{\alpha}{\alpha-1} \log\frac{\mathsf{P}_\alpha(X|B)_\rho}{ \mathsf{P}_\alpha(X)_\rho}, \quad \forall \alpha \in (1,\infty],
\end{equation}
which is a quantum generalization of the classical $\alpha$-leakage \cite{liao2019tunable}.
Moreover, we define the \textit{maximal $\alpha$-leakage} (Def.~\ref{def:max-alpha-Leakage}) to quantify how much information leaks about any sensitive function $S$ correlated to the non-sensitive $X$ via the released quantum system $B$:
\begin{equation}
    \mathcal{L}_\alpha^{\max} (X \to B)_\rho := \sup_{p_{S|X}:S-X-B} \mathcal{L}_\alpha (S \to B)_\rho, \quad \forall \alpha \in [1,\infty],
\end{equation}
where the maximization is over all stochastic transformation $p_{S|X}$ such that $S-X-B$ forms a classical-classical-quantum Markov chain.
The above quantity considers the maximal advantage of adversary for all sensitive data $S$ through a quantum privacy mechanism.

\subsection{Main Contributions}

In this paper, we characterize the above-mentioned quantities via certain quantum information measures.

First, we show that the maximal expected $\alpha$-gain is determined by the so-called \textit{measured conditional R\'enyi entropy of order $\alpha$}
(Theorem~\ref{theorem:alpha-gain}):
\begin{align} \label{eq:contribution-gain}
    \mathsf{P}_\alpha(X|B)_\rho = \mathrm{e}^{\frac{1-\alpha}{\alpha} H_\alpha^\mathds{M}(X|B)_\rho},
    \quad \forall\, \alpha \in [1,\infty].
\end{align}
The detailed definition of the measured conditional R\'enyi entropy $H_\alpha^\mathds{M}(X|B)_\rho$ with respect to a classical-quantum state $\rho_{XB}$ can be found in Def.~\ref{def:measured-condi-entropy} later.
This generalizes the classical result for characterizing the maximal expected $\alpha$-gain by the conditional R\'enyi entropy over classical privacy mechanisms \cite[Lemma~1]{liao2019tunable}.
Notably, for $\alpha = \infty$, our result recovers \textit{guessing probability}, which was shown by K{\"o}nig \textit{et al.}~\cite{Konig_2009} in 2009 as an operational meaning of min-entropy $H_{\infty}^\mathds{M}(X|B)_\rho$.
Hence, \eqref{eq:contribution-gain} is a parametric generalization of Ref.~\cite{Konig_2009}, and 
it provides an operational interpretation for measured conditional R\'enyi entropy $H_\alpha^\mathds{M}(X|B)_\rho$ for all $\alpha \in [1,\infty]$.

Second, we prove that $\alpha$-leakage is determined by the \textit{measured Arimoto information of order $\alpha$} (Theorem~\ref{thm:alpha_leakage}):
\begin{equation}
    \mathcal{L}_{\alpha} (X \rightarrow B)_\rho = I_{\alpha}^{\mathrm{A},\mathds{M}}(X:B)_\rho.
\end{equation}
The formal definition of measured Arimoto mutual information $I_{\alpha}^{\mathrm{A},\mathds{M}}$ is provided in Def.~\ref{def:measured-ArimotoMI} later.

Third, we prove that maximal $\alpha$-leakage is determined by the \textit{measured R\'enyi capacity} (Def.~\ref{def:measured_Renyi_capacity}), which is equal to both
\textit{measured Arimoto capacity} (Def.~\ref{def:measured_Arimoto_capacity}) and \textit{measured R\'enyi divergence radius} (Def.~\ref{def:div_radius}) for $\alpha \in (1,\infty]$, and by \textit{measured Arimoto information} (Def.~\ref{def:measured-ArimotoMI}) for $\alpha = 1$:
\begin{subnumcases}{\mathcal{L}_{\alpha}^{\max} (X \rightarrow B)_\rho= } 
    I_1^{\mathrm{A},\mathds{M}}(X:B)_\rho, & $\alpha=1$; \\
     C_{\alpha}^{\mathds{M}}(\mathscr{N}_{\textnormal{\texttt{supp}}(p_X) \to B}) = C_{\alpha}^{\mathrm{A},\mathds{M}}(\mathscr{N}_{\textnormal{\texttt{supp}}(p_X) \to B}) = R_{\alpha}^{\mathds{M}}(\mathscr{N}_{\textnormal{\texttt{supp}}(p_X) \to B}),  \label{eq:contribution-max-leakage->1}
     & $\alpha > 1$. 
\end{subnumcases}
Note here that for $\alpha>1$, the maximal $\alpha$-leakage depends on the input distribution only through its support.

Moreover, we present some interesting properties for maximal $\alpha$-leakage, such as data-processing inequality (DPI) for $\alpha \in [1,\infty]$ (Theorem~\ref{thm:properties}, part 4) and composition property (Theorem~\ref{thm:sub_Markov}). 
In the end, we discuss the asymptotic behavior of $\alpha$-leakage and maximal $\alpha$-leakage for $n$-fold product quantum privacy mechanisms $\mathscr{N}_{\mathcal{X}\to B}^{\otimes n}$.
We prove that, in the asymptotic limit $n\to \infty$, the maximal $\alpha$-leakage is characterized by the \emph{sandwiched R\'enyi divergence radius} (Theorem~\ref{thm:regularized_capacity}), i.e.
\begin{align} \label{eq:contribution-reg}
    \lim_{n\rightarrow\infty} \frac{1}{n} \mathcal{L}_{\alpha}^{\max} (X^n \rightarrow B^n)_{\rho^n} 
    =\lim_{n\rightarrow\infty} \frac{1}{n} C_\alpha^{\mathds{M}}(\mathscr{N}_{\mathcal{X}\to B}^{\otimes n} )
    = C^*_{\alpha}(\mathscr{N}_{\mathcal{X} \to B})
    = R^*_{\alpha}(\mathscr{N}_{\mathcal{X} \to B}), 
    \quad \forall\, \alpha>1.
\end{align}
Here, the underlying joint state $\rho_{X^n B^n}^n$
is induced by arbitrary fully-supported input distribution on non-sensitive data $X^n$ and $\mathscr{N}_{\mathcal{X}\to B}^{\otimes n}$.
%
The left most equality of \eqref{eq:contribution-reg} directly comes from \eqref{eq:contribution-max-leakage->1}.
For the second equality, we show that the regularized measured R\'enyi capacity equals sandwiched R\'enyi capacity, which may be of independent interest
(Proposition~\ref{prop:regularized_capacity}).

Our results hence provide operational meanings for measured Arimoto information, measured R\'enyi divergence radius, and sandwiched R\'enyi divergence radius.


\medskip
This paper is organized as follows.
Section~\ref{sec:pre} formally introduces measured quantities and their properties.
In Section~\ref{sec:infor}, we define $\alpha$-leakage and maximal $\alpha$-leakage for a quantum privacy mechanism, establish their equivalence to measured Arimoto information and measured R\'enyi divergence radius, and derive various properties for maximal $\alpha$-leakage.
In Section~\ref{sec:dis}, we discuss the implications of our results in privacy-utility tradeoffs and list some open problems.

For readability, we defer long proofs to Appendices~\ref{app:div_radius}--\ref{app:proof_regularized_capacity}.

\section{Preliminaries} \label{sec:pre}
\subsection{Notation}
Throughout this paper, we let $\mathcal{X}$ and $\mathcal{Y}$ be finite sets.
Let $\mathcal{H}_X$ be an $|\mathcal{X}|$-dimensional Hilbert space (i.e.~complex Euclidean space) with an orthonormal basis $\{| x \rangle\}_{x\in\mathcal{X}}$.
The outer product $|x\rangle\langle x|$ means an orthogonal projection onto the subspace spanned by vector $|x\rangle$.
The set of probability distributions on $\mathcal{X}$ is denoted as $\mathcal{P}(\mathcal{X})$.
For a probability distribution $p_X \in \mathcal{P}(\mathcal{X})$, we
denote its support by $\texttt{supp}(p_X)$.
Let $\mathcal{S}(\mathcal{H}_B)$ be the set of density matrices 
(i.e.~positive semi-definite matrices with unit trace)
on Hilbert space $\mathcal{H}_B$, and $\mathds{1}_B$ be the identity matrix on $\mathcal{H}_B$. 
The state of a quantum system $B$ is modeled by some density matrix $\rho_B \in \mathcal{S}(\mathcal{H}_B)$.
We denote $\texttt{supp}(\rho_B)$ as the subspace spanned by the set of vectors in the eigenspace of $\rho_B$ corresponding to positive eigenvalues.
For two Hermitian matrices $K$ and $L$ on the same Hilbert space, we define the Hilbert--Schmidt inner product as
$ \left\langle K, L\right\rangle := \Tr\left[ K L \right]$, 
where $\Tr$ is the standard trace. 

A classical-quantum (c-q) state on $\mathcal{S}(\mathcal{H}_X \otimes \mathcal{H}_B)$ is $\rho_{XB}:=\sum_{x\in\mathcal{X}} p_X(x)\ket{x}\!\!\bra{x}\otimes\rho_B^x$, where $p_X \in \mathcal{P}(\mathcal{X})$ and each $\rho_B^x \in \mathcal{S}(\mathcal{H}_B)$.
Namely, a c-q state $\rho_{XB}$ represents a quantum ensemble $\{ p_X (x), \rho_B^x\}_{x\in\mathcal{X}}$.
Let $\{\Pi_B^x\}_{x\in\mathcal{X}}$ be a POVM, i.e.~each $\Pi_B^x$ is a positive semi-definite matrix on $\mathcal{H}_B$ and $\sum_{x\in\mathcal{X}} \Pi_B^x = \mathds{1}_B$;
equivalently, we also use $\Pi_{XB}:=\sum_{x\in\mathcal{X}}\ket{x}\!\!\bra{x}\otimes\Pi_B^x$ to denote it for brevity. For a c-q state $\rho_{XB}$, the partial trace over system $X$ is denoted by $\Tr_X[\rho_{XB}] = (\Tr_X \otimes \mathrm{id}_B)(\rho_{XB}) = \sum_{x \in \mathcal{X} } (\bra{x} \otimes \mathds{1}_B) \rho_{XB} (\ket{x} \otimes \mathds{1}_B)$, where $\mathrm{id}_B$ is the identity map on system $B$.

For a power function $f: M \mapsto M^\alpha$ of a matrix $M$, we refer to $f$ as acting on the support of $M$.
Moreover, let $\mathscr{N}_{\mathcal{X} \to B }: \mathcal{X} \ni x \mapsto \rho_B^x \in \mathcal{S}(\mathcal{H}_B)$ be a c-q channel that maps each letter in the input set $x \in \mathcal{X}$ to a density matrix $\rho_B^x\in \mathcal{S}(\mathcal{H}_B)$.
The exponential function is denoted by $(\cdot)\mapsto \mathrm{e}^{(\cdot)}$.
The logarithmic function $(\cdot)\mapsto\log(\cdot)$ denotes natural logarithm.

\subsection{Information-Theoretic Quantities}
In this section, we briefly review the measured counterpart of entropic quantities, and refer readers to \cite{berta2017variational,hiai2017different} for more details.

\begin{definition}\label{def:big_def}
Let $\rho_{XB} = \sum_x p_X(x)\ket{x}\!\!\bra{x}\otimes\rho_B^x \in \mathcal{S}(\mathcal{H}_X\otimes \mathcal{H}_B)$ be a c-q state.
\begin{enumerate}
\item \textbf{Order-$\alpha$ R\'enyi divergence} \textnormal{\cite{Ren62}}: For $\alpha \in (0,1) \cup (1,\infty)$ and probability distributions $p_Y, q_Y \in \mathcal{P}(\mathcal{Y})$,
    \begin{equation}
    D_\alpha(p_Y \rVert q_Y) := \frac{1}{\alpha - 1} \log \sum_{y \in \mathcal{Y}} p_Y(y)^{\alpha} q_Y(y)^{1-\alpha},
    \end{equation}
    if $\alpha \in(0,1)$ or
    $\textnormal{\texttt{supp}}(p_Y) \subseteq \textnormal{\texttt{supp}}(q_Y)$; otherwise, it is defined as $+\infty$.
    In the limit $\alpha \to 1$, since R\'enyi divergence converges to Kullback--Leibler divergence, we denote it as $D_1(p_Y\rVert q_Y) \equiv D(p_Y\rVert q_Y) := \sum_{y\in\mathcal{Y}} p_Y(y) \log \frac{p_Y(y)}{q_Y(y)}$.
    As $\alpha \to \infty$, R\'enyi divergence converges to max-divergence
    \begin{equation}
        D_\infty(p_Y \rVert q_Y) :=
    \lim_{\alpha\to \infty} D_\alpha (p_Y\rVert q_Y) =
    \sup_{y\in \mathcal{Y}}\log\frac{p_Y(y)}{q_Y(y)}.
    \end{equation}
    The order-$0$ R\'enyi divergence is defined by taking limit $\alpha \to 0$.
    
\item \textbf{Measured R\'enyi divergence} \textnormal{\cite{berta2017variational}}: For $\alpha \in [0,\infty]$, density matrix $\rho$, and positive semi-definite matrix $\sigma$,
\begin{equation}
    D_\alpha^\mathds{M}( \rho \rVert \sigma) := \sup_{(\mathcal{Y}, \Pi)} D_\alpha\left( \{\Tr[\rho \Pi_y]\}_{y\in\mathcal{Y}} \rVert \{\Tr[\sigma \Pi_y]\}_{y\in\mathcal{Y}} \right).
\end{equation}
The supremum is over all finite sets $\mathcal{Y}$ and POVMs $\{\Pi^y\}_{y\in\mathcal{Y}}$. For $\alpha \in (0,\infty)\backslash\{1\}$, sometimes we also express measured R\'enyi divergence as:
\begin{equation}
    D_\alpha^\mathds{M} (\rho\rVert\sigma) = \frac{1}{\alpha - 1}\log Q_\alpha^\mathds{M}(\rho\rVert\sigma), 
\end{equation}
where \textbf{measured R\'enyi quasi divergence} is defined by
\begin{subnumcases}
    {Q_\alpha^\mathds{M}( \rho \rVert \sigma) := }
    \sup_{(\mathcal{Y}, \Pi)} \sum_{y \in \mathcal{Y}} (\Tr[\rho \Pi_y])^\alpha (\Tr[\sigma \Pi_y])^{1-\alpha}, & $\alpha \in (1,\infty)$\textnormal{;}\\
    \inf_{(\mathcal{Y}, \Pi)} \sum_{y \in \mathcal{Y}} (\Tr[\rho \Pi_y])^\alpha (\Tr[\sigma \Pi_y])^{1-\alpha}, & $\alpha \in (0,1)$.
\end{subnumcases}

\item \textbf{Measured conditional R\'enyi entropy} \textnormal{\cite[Eq.~(64)]{hanson2021guesswork}}: \label{def:measured-condi-entropy}
For $\alpha \in [0,\infty]$ and a c-q state $\rho_{XB}$,
\begin{equation}
    H^{\mathds{M}}_\alpha(X|B)_\rho := - \inf_{\sigma_B\in \mathcal{S}(\mathcal{H}_B)} D_\alpha^{\mathds{M}}(\rho_{XB}\rVert\mathds{1}_X\otimes\sigma_B).
\end{equation}

\item \textbf{Measured R\'enyi information} \textnormal{\cite{beigi2023lower}}: \label{def:measured_MI}
For $\alpha \in [0,\infty]$  and a c-q state $\rho_{XB}$,
\begin{equation}
    I_{\alpha}^{\mathds{M}}(X:B)_\rho := \inf_{\sigma_B\in \mathcal{S}(\mathcal{H}_B)}D_\alpha^{\mathds{M}}(\rho_{XB}\rVert\rho_X\otimes\sigma_B).
\end{equation}

\item \textbf{Measured Arimoto information} \textnormal{\cite{Ari75}}: \label{def:measured-ArimotoMI}
For $\alpha \in [0,\infty]$ and a c-q state $\rho_{XB}$,
\begin{equation}
    I_{\alpha}^{\mathrm{A},\mathds{M}}(X:B)_\rho := H_\alpha(X)_p - H^{\mathds{M}}_\alpha(X|B)_\rho,
\end{equation}
where
\begin{equation}
    H_\alpha(X)_p := \frac{1}{1-\alpha} \log \sum_{x \in \mathcal{X}} p_X(x)^\alpha,
\end{equation} and for $\alpha=1 \textit{ and }\infty$, $H_\alpha(X)_p$ is defined as taking limit $\alpha\rightarrow 1 \textit{ and }\alpha\rightarrow\infty$ respectively.

\item \textbf{Measured R\'enyi divergence radius} \textnormal{\cite{beigi2023lower}}: \label{def:div_radius}
For a c-q channel $\mathscr{N}_{\mathcal{X} \to B}: x\mapsto\rho_B^x$ and $\alpha \in [0,\infty]$,
\begin{equation}
    R_\alpha^{\mathds{M}}(\mathscr{N}_{\mathcal{X} \to B}) := \inf_{\sigma_B\in \mathcal{S}(\mathcal{H}_B)} \sup_{x\in\mathcal{X}} D_\alpha^{\mathds{M}}(\mathscr{N}(x)\rVert\sigma_B) \equiv \inf_{\sigma_B\in \mathcal{S}(\mathcal{H}_B)} \sup_{x\in\mathcal{X}} D_\alpha^{\mathds{M}}(\rho_B^x\rVert\sigma_B),
\end{equation}
which is the measured counterpart of the quantum R\'enyi
divergence radius \textnormal{\cite[Eq.~(85)]{mosonyi2017strong}}.

\item \textbf{Measured R\'enyi capacity} \textnormal{\cite{beigi2023lower}}: \label{def:measured_Renyi_capacity}
For $\alpha \in [0,\infty]$ and a c-q channel $\mathscr{N}_{\mathcal{X} \to B}: x\mapsto\rho_B^x$,
\begin{equation}
    C_{\alpha}^{\mathds{M}}(\mathscr{N}_{\mathcal{X} \to B}) := \sup_{p_X \in \mathcal{P}(\mathcal{X})} I_{\alpha}^{\mathds{M}}(X:B)_\rho.
\end{equation}

\item \textbf{Measured Arimoto capacity}: \label{def:measured_Arimoto_capacity}
For $\alpha \in [0,\infty]$ and a c-q channel $\mathscr{N}_{\mathcal{X} \to B}: x\mapsto\rho_B^x$,
\begin{equation}
    C_{\alpha}^{\mathrm{A},\mathds{M}}(\mathscr{N}_{\mathcal{X} \to B}) := \sup_{p_X \in \mathcal{P}(\mathcal{X})} I_{\alpha}^{\mathrm{A},\mathds{M}}(X:B)_\rho.
\end{equation}

\item \textbf{Sandwiched R\'enyi divergence} \textnormal{\cite{Martin2013quantumrenyi,mark2014strongconverse}}:
For density matrix $\rho$, positive semi-definite matrix $\sigma$, and $\alpha \in [0,1) \cup (1,\infty]$,
\begin{equation}
    D^*_{\alpha}\left(\rho\rVert\sigma \right) := \frac{1}{\alpha-1}\log \Tr \left[ \left(\sigma^{\frac{1-\alpha}{2\alpha}}\rho\sigma^{\frac{1-\alpha}{2\alpha}}\right)^{\alpha}\right]
\end{equation}
if $\alpha \in [0, 1)$ or $\textnormal{\texttt{supp}}(\rho) \subseteq \textnormal{\texttt{supp}}(\sigma) $; otherwise, it is defined as $+\infty$.

For $\alpha \in [\sfrac12,1) \cup (1,\infty]$, $D^*_{\alpha}$ satisfies data-processing inequality
\textnormal{\cite{Martin2013quantumrenyi, mark2014strongconverse,beigi2013sandwiched}}
.

\item \textbf{Umegaki's quantum relative entropy} \textnormal{\cite{umegaki1954conditional}}:
For density matrix $\rho$ and positive semi-definite matrix $\sigma$,
\begin{equation}
    D\left(\rho\rVert\sigma \right) := \Tr\left[\rho(\log\rho-\log\sigma)\right],
\end{equation}
if $\textnormal{\texttt{supp}}(\rho) \subseteq \textnormal{\texttt{supp}}(\sigma) $; otherwise, it is defined as $+\infty$.
Continuous extension of sandwiched R\'enyi divergence for $\alpha\to 1$ reduces to Umegaki's quantum relative entropy.

\item \textbf{Sandwiched R\'enyi information} \textnormal{\cite{CGH18, li2022operational,beigi2023lower}}:
For $\alpha \in [0,\infty]$ and a c-q state $\rho_{XB}$,
\begin{equation}
 I^*_\alpha\left(X : B\right)_{\rho} := \inf_{\sigma_B\in \mathcal{S}(\mathcal{H}_B)}D^*_{\alpha}(\rho_{XB}\rVert\rho_{X}\otimes\sigma_{B}).   
\end{equation}
\item \textbf{Sandwiched Arimoto information}:
For $\alpha \in [0,\infty]$ and a c-q state $\rho_{XB}$,
\begin{equation}
 I^{\mathrm{A},*}_\alpha\left(X : B\right)_{\rho} := H_\alpha(X)_p +\inf_{\sigma_B\in \mathcal{S}(\mathcal{H}_B)}D^*_{\alpha}(\rho_{XB}\rVert\mathds{1}_X\otimes\sigma_B).   
\end{equation}
\item \textbf{Sandwiched R\'enyi divergence radius} \textnormal{\cite[Eq.~(85)]{mosonyi2017strong}}: \label{def:sandwiched_div_radius}
For a c-q channel $\mathscr{N}_{\mathcal{X} \to B}: x\mapsto\rho_B^x$ and $\alpha \in [0,\infty]$,
\begin{equation}
    R_\alpha^*(\mathscr{N}_{\mathcal{X} \to B}) := \inf_{\sigma_B\in \mathcal{S}(\mathcal{H}_B)} \sup_{x\in\mathcal{X}} D_\alpha^*(\mathscr{N}(x)\rVert\sigma_B) \equiv \inf_{\sigma_B\in \mathcal{S}(\mathcal{H}_B)} \sup_{x\in\mathcal{X}} D_\alpha^*(\rho_B^x\rVert\sigma_B).
\end{equation}

\item \textbf{Sandwiched R\'enyi capacity} \textnormal{\cite{beigi2023lower}}:
For $\alpha \in [0,\infty]$ and a c-q state $\rho_{XB}$,
\begin{equation}
    C^*_{\alpha}(\mathscr{N}_{\mathcal{X} \to B}) := \sup_{p_X \in \mathcal{P}(\mathcal{X})} I^*_{\alpha}(X:B)_\rho.
\end{equation}

\item \textbf{Sandwiched Arimoto capacity}: \label{def:sandwiched_Arimoto_capacity}
For $\alpha \in [0,\infty]$ and a c-q channel $\mathscr{N}_{\mathcal{X} \to B}: x\mapsto\rho_B^x$,
\begin{equation}
    C_{\alpha}^{\mathrm{A},*}(\mathscr{N}_{\mathcal{X} \to B}) := \sup_{p_X \in \mathcal{P}(\mathcal{X})} I_{\alpha}^{\mathrm{A},*}(X:B)_\rho.
\end{equation}

\end{enumerate}
\end{definition}

For $\alpha \in \{\sfrac12, \infty\}$, measured R\'enyi divergence coincides with sandwiched R\'enyi divergence \cite{Konig_2009,berta2017variational,mosonyi2014quantum,fuchs1996distinguishability}.
In particular for $\alpha = +\infty$, it is named \emph{maximum relative entropy} \cite{Dat09}, i.e.~
\begin{align} \label{eq:max-relative-entropy}
D^{\mathds{M}}_{\infty} (\rho \rVert \sigma)
=D^*_\infty(\rho \rVert \sigma)
= \inf \{ \lambda \in\mathds{R}: \rho \leq \mathrm{e}^\lambda \sigma \}.
\end{align}    
For all $\alpha \in (\sfrac{1}{2}, \infty)$, we have the following strict relation \cite[Thm.~6]{berta2017variational}:
\begin{equation} \label{eq:mea_leq_san}
    D^\mathds{M}_\alpha(\rho \rVert\sigma) < D^*_\alpha(\rho\rVert\sigma),
\end{equation} 
unless $\rho$ commutes with $\sigma$.

\begin{remark}
    We remark that in the classical case (i.e.~$\{\rho_B^x\}_x$ mutually commute), the measured Arimoto information reduced to the classical Arimoto information introduced in Refs.~\cite{Ari75, Ver15}.
\end{remark}

\begin{definition} [$\alpha$-tilted distribution \cite{liao2019tunable,Ver15}] \label{def:tilted-alpha}
Given a parameter $\alpha \in (0,\infty)$ and a probability distribution $p_X \in \mathcal{P}(\mathcal{X})$, the $\alpha$-tilted distribution of $p_X$ is defined as
\begin{equation}
    p_X^{(\alpha)}(x) := \frac{p_X(x)^\alpha}{\sum_{x\in \mathcal{X}}p_X(x)^\alpha}.
\end{equation}
    
\end{definition}

\begin{lemma}[Variational formula of measured R\'enyi divergence {\cite[Lemma.~1,~3]{berta2017variational}, \cite{Hiai2021quantumf, CG24a}}] \label{lemma:var_measure}
For density matrices $\rho, \sigma$ and $\alpha \in [0, \infty]$,
\begin{subnumcases}
{D_{\alpha}^{\mathds{M}}(\rho \rVert \sigma)=}
    \sup_{\omega > 0} \Tr \left[ \rho\log \omega \right]-\log \Tr\left[ \sigma\omega\right] = \sup_{\omega > 0} \Tr \left[ \rho\log \omega \right] + 1 - \Tr\left[ \sigma\omega\right], & $\alpha = 1$\textnormal{;} \\
    \sup_{\omega > 0} \frac{1}{\alpha-1} \log \left(\left(\Tr\left[ \rho\omega^{1-\frac{1}{\alpha}} \right]\right)^\alpha \left(\Tr\left[ \sigma \omega\right]\right)^{1-\alpha}\right), & $\alpha \neq 1$. \label{eq:projectiveentropy}
    
\end{subnumcases}
\end{lemma}

\begin{lemma}[Data-processing inequality of measured R\'enyi divergence {\cite[Prop.~5.4]{Hiai2021quantumf}}] \label{lemma:DPI_measure}
Let $\rho, \sigma$ be density matrices and $\mathscr{N}$ be a fully quantum channel. Then, for all $\alpha \in [0,\infty]$,
\begin{equation}
     D_\alpha^\mathds{M}( \rho \rVert \sigma) \geq D_\alpha^\mathds{M}\left( \mathscr{N}(\rho) \rVert \mathscr{N}(\sigma)\right).
\end{equation}
\end{lemma}

\begin{lemma}[Super-additivity of measured R\'enyi divergence {\cite[Eq.~(65)]{berta2017variational}}]
\label{lemma:Super-additivity_measure}
Let $\rho_A, \sigma_A \in \mathcal{S}(\mathcal{H}_A)$ and $\rho_B, \sigma_B \in \mathcal{S}(\mathcal{H}_B)$ be density matrices. For all $\alpha \in [0,\infty]$,
\begin{equation}
     D_\alpha^\mathds{M}( \rho_A \otimes \rho_B \rVert \sigma_A \otimes \sigma_B) \geq D_\alpha^\mathds{M}\left( \rho_A \rVert \sigma_A\right)+D_\alpha^\mathds{M}\left( \rho_B \rVert \sigma_B\right).
\end{equation}
\end{lemma}

\begin{lemma}[Super-additivity of operational quantities {\cite[Sec.~V]{berta2017variational}}]
\label{lemma:Super-additivity_opr}
Let $\rho_A \in \mathcal{S}(\mathcal{H}_A)$ and $\rho_B \in \mathcal{S}(\mathcal{H}_B)$ be density matrices. 
Then, for all $\alpha \in [0,\infty]$,
\begin{align}
     &\inf_{ \sigma_{AB}\in\mathcal{S}(\mathcal{H}_A \otimes \mathcal{H}_B)} D_\alpha^\mathds{M}\left( \rho_A \otimes \rho_B \rVert \sigma_{AB}\right) 
     &\geq 
     \inf_{\sigma_A\in\mathcal{S}(\mathcal{H}_A)} D_\alpha^\mathds{M}\left( \rho_A \rVert \sigma_A\right) + \inf_{\sigma_B\in \mathcal{S}(\mathcal{H}_B)} D_\alpha^\mathds{M}\left( \rho_B \rVert \sigma_B\right).
\end{align}
\end{lemma}

\begin{lemma}[Invariant maximal value over $\alpha$-tilted distribution {\cite[Prop.~1]{Kam2024newAlgArimoto}}] \label{lemma:alpha-tilted}
    Let $\alpha \in (0,\infty)$. Given a distribution $p_X \in \mathcal{P}(\mathcal{X})$ and a continuous function $f: \mathcal{P}(\mathcal{X}) \to \mathbb{R}$,
    \begin{equation}
        \max_{p_X \in \mathcal{P}(\mathcal{X})} f(p_X) = \max_{p_X \in \mathcal{P}(\mathcal{X})} f\left(p_X^{(\alpha)}\right).
    \end{equation}
\end{lemma}

\begin{lemma}[Equivalent expressions of R\'enyi capacity, R\'enyi divergence radius, and Arimoto capacity] \label{lemma:div_radius}
For $\alpha \geq 0$ and a c-q channel $\mathscr{N}_{\mathcal{X} \to B}: x\mapsto\rho_B^x$ , we have
\begin{equation}
    C_{\alpha}^{\mathds{M}}(\mathscr{N}_{\mathcal{X} \to B})  = R_\alpha^{\mathds{M}}(\mathscr{N}_{\mathcal{X} \to B}) = C_{\alpha}^{\mathrm{A},\mathds{M}}(\mathscr{N}_{\mathcal{X} \to B});
\end{equation}
on the other hand, for $\alpha \geq \sfrac{1}{2}$, 
\begin{equation}
    C_{\alpha}^*(\mathscr{N}_{\mathcal{X} \to B})  = R_\alpha^*(\mathscr{N}_{\mathcal{X} \to B}) = C_{\alpha}^{\mathrm{A},*}(\mathscr{N}_{\mathcal{X} \to B}).
\end{equation}
\end{lemma}
Beigi and Tomamichel \cite[Lemma 1]{beigi2023lower} proved the equivalence of measured Rényi capacity and measured R\'enyi divergence radius for $\alpha \in (0,1)$ and stated that the proof for $\alpha \geq 1$ follows similarly. For completeness, we provide the proof for the $\alpha \geq 0$ case in Appendix~\ref{app:div_radius} and further prove their equivalence to measured Arimoto capacity.

On the other hand, the equivalence of sandwiched Rényi capacity and sandwiched R\'enyi divergence radius for $\alpha \geq \sfrac{1}{2}$ is proved by Mosonyi and Ogawa \cite[Prop.~4.2]{mosonyi2017strong}. For completeness, we also prove their equivalence to sandwiched Arimoto capacity in Appendix~\ref{app:div_radius}.

\section{Information Leakage Measures and Main Results} \label{sec:infor}
In this section, we first introduce the notions of the  \textit{expected $\alpha$-loss} and \textit{expected $\alpha$-gain}, characterizing how much information one loses or gains from observing a quantum ensemble.
Subsequently, we propose the definitions of $\alpha$-leakage and maximal $\alpha$-leakage as measures of information leakage under a quantum privacy mechanism. 
These measures capture how much inference an adversary can draw when the data are released via a privacy mechanism.
The purpose for a quantum adversary is to minimize expected $\alpha$-loss, which is equivalent to maximize expected $\alpha$-gain.\footnote{
The proposed \textit{expected $\alpha$-loss} and \textit{expected $\alpha$-gain} recovers the classical correspondence in the setting of classical privacy mechanisms \cite{liao2019tunable, diaz2019robustness}.
Instead of characterizing $\alpha$-leakage by expected $\alpha$-loss as in Ref.~\cite{liao2019tunable}, we depict $\alpha$-leakage via expected $\alpha$-gain so as to obtain more insight.}

We further prove that $\alpha$-leakage is determined by  measured Arimoto information and that maximal $\alpha$-leakage is determined by measured R\'enyi capacity with a support constraint respectively.
We also prove that regularized $\alpha$-leakage and regularized maximal $\alpha$-leakage for i.i.d.~quantum privacy mechanisms are equivalent to sandwiched information and sandwiched capacity, respectively.

\subsection{The Expected \texorpdfstring{$\alpha$}{alpha}-Loss and \texorpdfstring{$\alpha$}{alpha}-Gain}

Focusing on the study of information leakage for quantum privacy mechanisms, we mainly discuss the range $\alpha \geq 1$ of $\alpha$-loss.
In fact, the range $\alpha \in (0,1)$ of $\alpha$-loss was also defined in classical literature (see e.g.~\cite{sypherd2022alphaloss}). We define  $\alpha$-loss for $\alpha < 1$ in Appendix~\ref{app:alpha-loss<1}.

\begin{definition}[Expected $\alpha$-loss]
Consider a c-q state $\rho_{XB} =\sum_x p_X(x)\ket{x}\!\!\bra{x}\otimes\rho_B^x$.
For any $\alpha \in [1,\infty]$, we define an \textit{expected $\alpha$-loss} of POVM $\Pi_{XB}$ under $\rho_{XB}$ as:

$\varepsilon_\alpha (\Pi_{XB})_\rho 
    := $
\begin{subnumcases}
    { }
    \frac{\alpha}{\alpha-1}\left(1 - \Tr\left[ \rho_{XB} \Pi_{XB}^{\frac{\alpha-1}{\alpha}} \right]\right), & if $\alpha > 1$ or $\textnormal{\texttt{supp}}(\rho_{XB}) \subseteq \textnormal{\texttt{supp}}(\Pi_{XB})$;\label{eq:expected-loss}\\
    -\Tr\left[ \rho_{XB} \log\Pi_{XB} \right], & if $\alpha = 1$ and $\textnormal{\texttt{supp}}(\rho_{XB}) \subseteq \textnormal{\texttt{supp}}(\Pi_{XB})$;\\
    +\infty, & otherwise.
\end{subnumcases}

\end{definition}

\begin{remark}
    Note that expected $\alpha$-loss defined in \eqref{eq:expected-loss} is always non-negative because $\Pi_{XB} \leq \Pi_{XB}^{\frac{\alpha-1}{\alpha}}
    \leq
    \mathds{1}_{XB}$ for $\alpha \in [1,\infty]$.
\end{remark}

\begin{remark}
    For $\alpha = 1$, $\varepsilon_1 (\Pi_{XB})_\rho = -\Tr\left[ \rho_{XB} \log\Pi_{XB} \right]$ is known as \textit{expected log-loss} \cite{liao2019tunable};
    as for $\alpha = \infty$, 
    \begin{align}
    \varepsilon_{\infty} (\Pi_{XB})_\rho 
    = 1 - \sum_{x\in\mathcal{X}} p_X(x)\Tr\left[ \rho_B^x \Pi_B^x \right]
    \end{align}
    corresponds to the error probability of discriminating the ensemble $\{p_X(x), \rho_B^x\}_{x\in\mathcal{X}}$ using the POVM $\{\Pi_B^x\}_{x\in\mathcal{X}}$ (see e.g.~\cite[Chapter 3]{Wat18}).
    Therefore, the expected $\alpha$-loss $\varepsilon_\alpha $ serves as a tunable loss measure that interpolates between the various known loss (or risk) functions.
\end{remark}

\begin{remark}
 One may introduce an \emph{$\alpha$-loss operator} of a POVM $\Pi_{XB}$ defined as
 \begin{subnumcases}
    {\ell_\alpha(\Pi_{XB}) := }\frac{\alpha}{\alpha-1}\left(\mathds{1}_{XB} - \Pi_{XB}^{\frac{\alpha-1}{\alpha}}\right), & if $\alpha > 1$ or $\Pi_{XB} > 0$;\\
    -\log \Pi_{XB}, & if $\alpha = 1$ and $\Pi_{XB} > 0$;\\
    +\infty, & otherwise,
\end{subnumcases}
such that $\varepsilon_\alpha(\Pi_{XB})_\rho = \langle \rho_{XB}, \ell_\alpha(\Pi_{XB})\rangle$.
In the classical setting, such an $\alpha$-loss operator is called $\alpha$-loss function in \cite[Def.~3]{liao2019tunable}.
\end{remark}

\begin{definition}[Minimal expected $\alpha$-loss]
    For any c-q state
    $\rho_{XB}=\sum_x p_X(x)\ket{x}\!\!\bra{x}\otimes\rho_B^x$ and $\alpha \in [1,\infty]$,
    we define the \textit{minimal expected $\alpha$-loss} for $\rho_{XB}$ as:
    \begin{subnumcases}
    {\varepsilon_\alpha(X|B)_\rho := \inf_{\textnormal{POVM}\, \Pi_{XB}} \varepsilon_\alpha (\Pi_{XB})_\rho
    = }    
    \inf_{\textnormal{POVM}\, \Pi_{XB}} \frac{\alpha}{\alpha-1}\left(1 - \Tr\left[ \rho_{XB} \Pi_{XB}^{\frac{\alpha-1}{\alpha}} \right]\right),&  for $\alpha > 1$;\\
    \inf_{\textnormal{POVM}\, \Pi_{XB}}  - \Tr\left[ \rho_{XB}\log\Pi_{XB} \right], & for $\alpha = 1$,
    \end{subnumcases}
    where the minimization is over all \textnormal{POVM}s on $\mathcal{H}_B$, i.e.~for all POVM $\Pi_{XB}$ such that\\ $\Tr_X[\Pi_{XB}]= \mathds{1}_B$ and  $\Pi_{XB} \geq 0$.
    
    On the other hand, we define the \textit{minimal expected $\alpha$-loss} for an inference $\hat{X}$ without observing $B$ as
    \begin{subnumcases}
    {\varepsilon_\alpha(X)_\rho := }       \inf_{p_{\hat{X}}\in\mathcal{P}(\mathcal{X})} \frac{\alpha}{\alpha-1}\left(1 - \mathds{E}_{x\sim p_X}\left[ {p_{\hat{X}}}(x)^{\frac{\alpha-1}{\alpha}} \right]\right),&  for $\alpha > 1$;\\
    \inf_{p_{\hat{X}}\in\mathcal{P}(\mathcal{X})}  - \mathds{E}_{x\sim p_X}\left[ \log p_{\hat{X}}(x)\right], & for $\alpha = 1$.
    \end{subnumcases}
\end{definition}

\begin{remark}
    A trivial guess $\Pi_{XB} = \sum_{x\in\mathcal{X}} |x\rangle\langle x| \otimes \frac{\mathds{1}_B}{|\mathcal{X}|}$ in $\varepsilon_\alpha(X|B)_\rho$ and $p_{\hat{X}}(x) = \frac{1}{|\mathcal{X}|}$ in $\varepsilon_\alpha(X)_\rho$ provides the following upper bound on the minimal expected $\alpha$-loss:
    \begin{align}
    \varepsilon_\alpha(X|B)_\rho
    &\leq \frac{\alpha}{\alpha-1} \left( 1 - \left|\mathcal{X}\right|^{ \frac{1-\alpha}{\alpha} } \right)
    \\
    \varepsilon_\alpha(X)_\rho 
    &\leq 
    \frac{\alpha}{\alpha-1} \left( 1 - \left|\mathcal{X}\right|^{ \frac{1-\alpha}{\alpha} } \right).
    \end{align}
    Later in Theorem~\ref{theorem:alpha-gain}, we will provide explicit characterization of the minimal expected $\alpha$-loss.
\end{remark}

\begin{definition}[Expected $\alpha$-gain] \label{def:alpha-gain}
    Given a c-q state $\rho_{XB}=\sum_x p_X(x)\ket{x}\!\!\bra{x}\otimes\rho_B^x$ and $\alpha \geq1$, for POVM $\Pi_{XB}$,
    we define \textit{expected $\alpha$-gain} for $\rho_{XB}$ as:
    \begin{equation}
        g_\alpha(\Pi_{XB})_\rho :=  \Tr\left[ \rho_{XB} \Pi_{XB}^{\frac{\alpha-1}{\alpha}}\right].\\
    \end{equation}
\end{definition}

\begin{definition}[Maximal expected $\alpha$-gain] \label{def:max-alpha-gain}
    For any c-q state
    $\rho_{XB}=\sum_x p_X(x)\ket{x}\!\!\bra{x}\otimes\rho_B^x$, POVM $\Pi_{XB}$, and $\alpha \in [1,\infty]$,
    we define the \textit{maximal expected $\alpha$-gain} for $\rho_{XB}$ as:
    \begin{align}
        \mathsf{P}_\alpha(X |B)_\rho &:= \sup_{\textnormal{POVM}\, \Pi_{XB}} g_\alpha(\Pi_{XB})_\rho 
        = \sup_{\textsc{POVM}\, \{\Pi_B^x\}_{x\in\mathcal{X}}}\sum_{x\in\mathcal{X}} p_X(x)\Tr\left[ \rho_B^x (\Pi_B^x)^\frac{\alpha-1}{\alpha} \right],
    \end{align}
    where the maximization is over all \textnormal{POVMs} on $\mathcal{H}_B$, i.e.~for all $\Pi_{XB}$ such that $\Tr_X[\Pi_{XB}]= \mathds{1}_B$ and  $\Pi_{XB} \geq 0$.

    On the other hand, we define the \textit{maximal expected $\alpha$-gain} for an inference $\hat{X}$ without observing $B$ as
    \begin{equation}
        \mathsf{P}_\alpha(X)_\rho := \sup_{p_{\hat{X}}:\hat{X} \perp X} \mathds{E}_{x\sim p_X}\left[ \Pr(\hat{X} = X | X = x)^{\frac{\alpha-1}{\alpha}} \right] = \sup_{p_{\hat{X}}} \mathds{E}_{x\sim p_X}\left[ {p_{\hat{X}}}(x)^{\frac{\alpha-1}{\alpha}} \right],
    \end{equation}
    where the condition $\hat{X} \perp X$ of the supremum in the first expression springs from conditional independence of Markov chain $X-B-\hat{X}$. Without the prior knowledge embedded in $B$, an adversary can only guess $\hat{X}$ independently.
\end{definition}

\begin{remark}
For $\alpha = \infty$, the quantity $\mathsf{P}_\infty(X|B)_\rho$ is equal to the maximal success (guessing) probability of discriminating the ensemble $\{p_X(x), \rho_B^x\}_{x\in\mathcal{X}}$. 
Moreover, it is straightforward to calculate the maximal success (guessing) probability without observing the system $B$, i.e.~$\mathsf{P}_\infty(X)_\rho = \max_{x\in\mathcal{X}} p_X(x)$, placing all weights on the symbol with the maximal prior probability.
\end{remark}

\begin{remark}
    Note that for POVM $\Pi_{XB}>0$, minimal expected $\alpha$-loss and maximal expected $\alpha$-gain are related by
\begin{equation} \label{eq:relation}
    \mathsf{P}_\alpha(X|B)_\rho = 1-\frac{\alpha-1}{\alpha} \varepsilon_\alpha (X|B)_\rho.
\end{equation}
Moreover, maximal expected $\alpha$-gain lies in the interval $[0,1]$ because $\Pi_{XB} \leq \Pi_{XB}^{\frac{\alpha-1}{\alpha}} \leq \mathds{1}_{XB}$ for $\alpha \in [1,\infty]$.

\end{remark}

Our first result is the following characterization of the maximal expected $\alpha$-gain for a c-q state $\rho_{XB}$ via the measured conditional R\'enyi entropy.
This thereby provides an operational interpretation for measured conditional R\'enyi entropy.

\begin{theorem}[Characterization of the maximal expected $\alpha$-gain]\label{theorem:alpha-gain}
    For any classical-quantum state $\rho_{XB}$ and 
    $\alpha \in [1,\infty]$,
    the maximal expected $\alpha$-gain in Def.~\ref{def:max-alpha-gain} is given by
    \begin{equation} \label{eq:conj}
    \mathsf{P}_\alpha(X|B)_\rho = \mathrm{e}^{\frac{1-\alpha}{\alpha} H_\alpha^\mathds{M}(X|B)_\rho },
    \end{equation}
    where $H_\alpha^\mathds{M}(X|B)_\rho$ was introduced in Def. 1.~\ref{def:measured-condi-entropy}.
\end{theorem}
\noindent A detail proof for Theorem~\ref{theorem:alpha-gain} is provided in Appendix~\ref{app:alpha_gain}.

\begin{remark}
    For $\alpha = \infty$ and recalling the max relative entropy in 
    \eqref{eq:max-relative-entropy}, 
    the error exponent in \eqref{eq:conj} is reduced to the so-called min-entropy introduced by K{\"o}nig, Renner, and Schaffner \cite[Thm.~1]{Konig_2009}, i.e.,
    \begin{align}\label{eq:guess}
         -\log \mathsf{P}_\infty(X|B)_\rho &= {H^*_\infty(X|B)_\rho}
         \\
    &= -\inf_{\sigma_B \in \mathcal{S}(\mathcal{H}_B)} D^*_\infty(\rho_{XB} \parallel \mathds{1}_X\otimes\sigma_B)\\
    &= -\inf_{\sigma_B \in \mathcal{S}(\mathcal{H}_B)} 
 \inf \{ \lambda \in \mathds{R}: \rho_{XB} \leq \mathrm{e}^\lambda (\mathds{1}_X\otimes\sigma_B) \}.
    \end{align}
\end{remark}

Via \eqref{eq:relation}, the minimal expected $\alpha$-loss is expressed as 

\begin{subnumcases} 
    {\varepsilon_\alpha(X|B)_\rho =}
    \frac{\alpha}{\alpha-1}\left(1 - \mathrm{e}^{\frac{1-\alpha}{\alpha}H_\alpha^\mathds{M}(X|B)_\rho} \right), &for $\alpha >1$,
    \label{eq:alpha-loss-a}
    \\
    H^\mathds{M}_1(X|B)_\rho, &for $\alpha = 1$. 
    \label{eq:alpha-loss-b}
\end{subnumcases}

We remark that \eqref{eq:alpha-loss-a} and \eqref{eq:alpha-loss-b} recover the classical results given in \cite[Lemma 1]{liao2019tunable} and \cite[Proposition 1]{diaz2019robustness}.

\subsection{Main Results: \texorpdfstring{$\alpha$}{alpha}-Leakage and Maximal \texorpdfstring{$\alpha$}{alpha}-Leakage}

Let $\mathscr{N}_{\mathcal{X}\to B}: x \mapsto \rho_B^x$ denote a classical-quantum privacy mechanism. To quantify the multiplicative increase in the maximal expected gain of data $X$ when observing its released version $B$ via quantum privacy mechanisms, we introduce the following information-leakage measures: $\alpha$-leakage and maximal $\alpha$-leakage.

\begin{definition}[$\alpha$-leakage] \label{def:alpha-Leakage}
    Given $\alpha \in [1,\infty]$ and a c-q state $\rho_{XB}$, the $\alpha$-leakage from $X$ to $B$ is defined for $\alpha > 1$ as
\begin{align}        
    {\mathcal{L}}_{\alpha} (X \rightarrow B)_\rho := \frac{\alpha}{\alpha-1} \log\frac{\mathsf{P}_\alpha(X|B)_\rho}{ \mathsf{P}_\alpha(X)_\rho}.\label{eq:alpha-Leakage}    
\end{align}
When $\alpha = 1$, the $\alpha$-leakage  is 
\begin{align}
    {\mathcal{L}}_{1} (X \rightarrow B)_\rho    
    := \lim_{\alpha\to 1} {\mathcal{L}}_{\alpha} (X \rightarrow B)_\rho
    =
    \varepsilon_1(X)_\rho-\varepsilon_1(X|B)_\rho
\end{align}
by continuous extension.
\end{definition}

\begin{remark}
    Def.~\ref{def:alpha-Leakage} is an extension of classical $\alpha$-leakage introduced by Liao \textit{et al.}~\cite[Def.~5]{liao2019tunable} to the scenarios of quantum privacy mechanisms.
    Note that the denominator of the logarithmic term in \eqref{eq:alpha-Leakage} is maximal expected gain of a decision without additional information apart from $X$, while the numerator of the logarithmic term in \eqref{eq:alpha-Leakage} is maximal expected gain of a decision having access to quantum system $B$ \cite{diaz2019robustness}. In particular, for the case of $\alpha = \infty$, 
    the proposed $\alpha$-leakage recovers the correctly-guessing information leakage that Asoodeh \textit{et al.}~\cite{asoodeh2018estimation} proposed:
    \begin{equation}
        \mathcal{L}_{\infty}(X\to B)_\rho := \log \frac{\mathsf{P}_{\infty}(X|B)_\rho}{\mathsf{P}_{\infty}(X)_\rho}.
    \end{equation}
\end{remark}

\begin{theorem}[Characterization of $\alpha$-leakage] \label{thm:alpha_leakage}
    For $\alpha \in [1,\infty]$, $\alpha$-leakage defined in Def.~\ref{def:alpha-Leakage} can be expressed as
    \begin{equation} \label{eq:thm 1}
        \mathcal{L}_{\alpha} (X \rightarrow B)_\rho = I_{\alpha}^{\mathrm{A},\mathds{M}}(X:B)_\rho,
    \end{equation}
    where $I_{\alpha}^{\mathrm{A},\mathds{M}}(X:B)_\rho$ was introduced in
    Def. 1.~\ref{def:measured-ArimotoMI}.
\end{theorem}
\noindent A detailed proof of Theorem~\ref{thm:alpha_leakage} is provided in Appendix~\ref{app:alpha_leakage}. 

\medskip
In addition to capturing how much an adversary learns about $X$ from $B$, in practice, we are often more interested in quantifying how much information leaks via $B$ for any function $S$ of $X$. With this goal, now we introduce the definition of maximal $\alpha$-leakage.

\begin{definition}[Maximal $\alpha$-leakage] \label{def:max-alpha-Leakage}
Given a joint c-q state $\rho_{XB}\in\mathcal{S}(\mathcal{H}_X \otimes \mathcal{H}_B)$, the maximal $\alpha$-leakage from $X$ to $B$ is defined as
\begin{equation}
    \mathcal{L}_\alpha^{\max} (X \to B)_\rho := \sup_{p_{S|X}:S-X-B} \mathcal{L}_\alpha (S \to B)_\rho
\end{equation}
for $\alpha \in [1,\infty]$, where $S$ denotes any function of $X$ and takes values from an arbitrary finite set.

\end{definition}

\begin{theorem}[Characterization of the maximal $\alpha$-leakage] \label{thm:max_alpha_leakage}
    For $\alpha \in [1,\infty]$, the maximal $\alpha$-leakage defined in Def.~\ref{def:max-alpha-Leakage} can be expressed as \label{eq:thm 2}
        \begin{subnumcases}{\mathcal{L}_{\alpha}^{\max} (X \rightarrow B)_\rho= } 
        I_1^{\mathrm{A},\mathds{M}}(X:B)_\rho = {\mathcal{L}}_1 (X \rightarrow B)_\rho, & $\alpha = 1$\textnormal{;}\\
        C_{\alpha}^{\mathds{M}}(\mathscr{N}_{\textnormal{\texttt{supp}}(p_X) \to B}) = C_{\alpha}^{\mathrm{A},\mathds{M}}(\mathscr{N}_{\textnormal{\texttt{supp}}(p_X) \to B}) = R_{\alpha}^{\mathds{M}}(\mathscr{N}_{\textnormal{\texttt{supp}}(p_X) \to B}),  
        & $\alpha > 1$. \label{eq:max_alpha_bounded_C}
        \end{subnumcases}
\end{theorem}
\noindent A detailed proof of Theorem~\ref{thm:max_alpha_leakage} is in Appendix~\ref{app:max_alpha_leakage}.

When $\alpha=1$, maximal $\alpha$-leakage reduces to measured Arimoto mutual information $I_1^{\mathrm{A},\mathds{M}}(X:B)_\rho$; when $\alpha=\infty$, maximal $\alpha$-leakage reduces to maximal leakage proposed in 
\textnormal{\cite{farokhi2023maximal}}.

\begin{remark}
    Theorem~\ref{thm:max_alpha_leakage} provides an operational meaning of measured R\'enyi capacity. Note that when $\alpha = 1$, maximal $\alpha$-leakage depends on input probability distribution $p_X$; when $\alpha > 1$, maximal $\alpha$-leakage depends on input probability distribution only through its support $\textnormal{\texttt{supp}}(p_X)$.
    
    As stated in Lemma~\ref{lemma:div_radius}, 
    maximal $\alpha$-leakage is also equal to measured Arimoto capacity and measured R\'enyi divergence radius.
\end{remark}

\subsection{Properties of Maximal \texorpdfstring{$\alpha$}{alpha}-Leakage}

To further analyze the performance of quantum privacy mechanisms, now we explore some properties of maximal $\alpha$-leakage.

\begin{theorem} \label{thm:properties}
Denote by ${\bar{\rho}}_{XB} = \sum_{x \in \mathcal{X}} \bar{p}_{X}(x) |x\rangle \langle x| \otimes \rho_{B}^{x}$
for any input distribution $\bar{p}_{X} \in \mathcal{P}(\mathcal{X})$ as an optimization variable.
For $\alpha \in [1,\infty]$
and given a c-q state $\rho_{XB} = \sum_{x\in\mathcal{X}} p_X(x)|x\rangle \langle x|\otimes \rho_B^x$, maximal $\alpha$-leakage 
\begin{subnumcases}{\mathcal{L}_{\alpha}^{\max} (X \rightarrow B)_\rho= } 
    I_1^{\mathrm{A},\mathds{M}}(X:B)_\rho, & $\alpha = 1$\textnormal{;}\\
    \sup_{\bar{p}_X \in \mathcal{P}(\textnormal{\texttt{supp}}(p_X))} \inf_{\sigma_B\in \mathcal{S}(\mathcal{H}_B)} D_\alpha^{\mathds{M}}(\bar{\rho}_{XB}\rVert \bar{\rho}_X\otimes\sigma_B),  
    & $\alpha > 1$. 
\end{subnumcases}
has the following properties:
\begin{enumerate}
    \item is a concave program of optimization variable $\bar{p}_X$ for $\alpha > 1$, and a concave function of input distribution $p_X$ for $\alpha = 1$;
    \item is quasi-convex in $\rho_B^x$ given optimization variable $\bar{p}_X$ for $\alpha > 1$ or given input distribution $p_X$ for $\alpha = 1$;
    \item is non-decreasing in $\alpha$;
    \item satisfies data-processing inequality;
    \item 
\begin{subnumcases}
    {0 \leq \mathcal{L}_\alpha^{\max} (X \to B)_\rho \leq} \log(|\textnormal{\texttt{supp}}(p_X)|), & for $\alpha > 1$\textnormal{;}\\
    H_1(X)_p, & for $\alpha = 1$\textnormal{;}
\end{subnumcases}
\end{enumerate}
\end{theorem}
\noindent A detailed proof of Theorem~\ref{thm:properties} is provided in Appendix~\ref{app:properties}.

\medskip
Sometimes, an adversary can access more than one released version of non-sensitive data $X$. The following theorem shows that even an adversary receives multiple independently released data $B$, they cannot obtain information about sensitive data $S$ more than marginal sums of maximal $\alpha$-leakage. This behavior is also known as composition property \cite[Sec.~3]{kairouz2015composition} \cite[Lemma.~6]{issa2019operational} \cite[Thm.~5]{liao2019tunable}.

\begin{theorem} [Composition property] \label{thm:sub_Markov}
    Given a probability distribution $p_X \in \mathcal{P}(\mathcal{X})$ and quantum privacy mechanisms $\mathscr{N}_{\mathcal{X} \to B_1}:x\mapsto\rho_{B_1}^x$ and $\mathscr{N}_{\mathcal{X}\to B_2}:x\mapsto\rho_{B_2}^x$, for any $\alpha \in [1,\infty]$, the maximal $\alpha$-leakage from $X$ to $B_1 B_2$ is bounded above by 
\begin{equation}
    \mathcal{L}_\alpha ^{\max} (X \to B_1,B_2)_\rho \leq \mathcal{L}_\alpha ^{\max} (X \to B_1)_\rho + \mathcal{L}_\alpha ^{\max} (X \to B_2)_\rho,
\end{equation}
where $\rho_{XB_1B_2}:= \sum_{x\in\mathcal{X}}p_X(x) \ket{x}\!\!\bra{x} \otimes \rho_{B_1}^x \otimes \rho_{B_2}^x$.
\end{theorem}
\noindent A detailed proof of Theorem~\ref{thm:sub_Markov} is provided in Appendix~\ref{app:sub_Markov}.

\subsection{Asymptotic behaviors of \texorpdfstring{$\alpha$}{alpha}-leakage and maximal \texorpdfstring{$\alpha$}{alpha}-leakage}

In previous sections, we have considered $\alpha$-leakage and maximal $\alpha$-leakage for a quantum privacy mechanism in the \emph{one-shot setting};
namely, the underlying data $S$ and $X$ and the quantum privacy mechanism $\mathscr{N}_{X\to B}$ are used only once.
Here, we discuss the asymptotic behaviors of $\alpha$-leakage and maximal $\alpha$-leakage when non-sensitive data $X^n$ are released via i.i.d.~quantum privacy mechanisms:
\begin{align}
\mathscr{N}_{\mathcal{X}\rightarrow B}^{\otimes n}: 
x_1x_2\cdots x_n \mapsto
\rho_{B_1}^{x_1} \otimes 
\rho_{B_2}^{x_2} \otimes \cdots \otimes \rho_{B_n}^{x_n} =: \rho_{B^n}^{x^n}.
\end{align}
More precisely, we will study both $\mathcal{L}_{\alpha}$ and $\mathcal{L}_{\alpha}^{\max}$ under $\mathscr{N}_{\mathcal{X}\rightarrow B}^{\otimes n}$ and normalize the quantities by $n$ to study the average information leakage when $n\to \infty$, which is termed \emph{regularization} in quantum information theory.

While the i.i.d.~assumption is not often adopted in information-theoretic security because one cannot restrict quantum adversaries to attack only in an i.i.d.~manner;
nevertheless, in the scenario of information leakage, the quantum privacy mechanism is employed by the system designer, and an i.i.d.~quantum privacy mechanism $\mathscr{N}_{\mathcal{X}\rightarrow B}^{\otimes n}$ is arguably easier to perform than a general $n$-shot privacy mechanism $\mathscr{N}_{\mathcal{X}^n \rightarrow B^n}:$ that may output a multipartite entangled state on system $B^n$.

When non-sensitive data $X^n$ are i.i.d. as well, the following theorem shows that the asymptotic behavior of $\alpha$-leakage is characterized by sandwiched R\'enyi information but under the $\alpha$-tilted distribution 
$p_X^{(\alpha)}$.
Note that the one-shot $\alpha$-leakage $\mathcal{L}_{\alpha}(X \to B)_{\rho}$ is characterized by measured Arimoto information  $I^{\mathrm{A},\mathds{M}}_{\alpha}(X:B)_\rho$ (Theorem~\ref{thm:alpha_leakage}), which is in general not identical to 
measured R\'enyi information $I^{\mathds{M}}_{\alpha}(X:B)_\rho$
nor 
sandwiched R\'enyi information $I^{*}_{\alpha}(X:B)_\rho$.
Hence, the regularization of $\mathcal{L}_{\alpha}$ converges to a different quantity as expected (Theorem~\ref{thm:regularizedMI}).
The interested readers are referred to Lemmas~\ref{lemma:regularized_renyi_sandwiched} and \ref{lemma:regularized_arimoto_sandwiched} in Appendix~\ref{app:REG} for more details.

\begin{theorem}[Characterization of regularized $\alpha$-leakage] \label{thm:regularizedMI}
Let $\rho_{XB} = \sum_{x\in\mathcal{X}} p_X(x)|x\rangle \langle x|\otimes \rho_B^x$. For $\alpha \geq 1$, then regularized $\alpha$-leakage from i.i.d.~$X^n$ to $B^n$ under $\mathscr{N}_{\mathcal{X}\rightarrow B}^{\otimes n}$
is given by
\begin{equation}
  \lim_{n\rightarrow\infty} \frac{1}{n}\mathcal{L}_{\alpha}(X^{n} \rightarrow B^{n})_{\rho^{\otimes n}} = I^*_\alpha\left(X : B\right)_{\rho^{(\alpha)}},
\end{equation}
where the $\alpha$-tilted distribution 
$p_X^{(\alpha)}$ is introduced in Def.~\ref{def:tilted-alpha} and we denote
\begin{align}
    \label{eq:asym_Xbar}
    \rho^{(\alpha)}_{XB} &\equiv \sum_{x\in\mathcal{X}} p_X^{(\alpha)}(x)\ket{x}\!\!\bra{x} \otimes \rho_B^x.
\end{align}
\end{theorem}
\noindent 
Theorem~\ref{thm:regularizedMI} follows from the fact that regularized measured R\'enyi divergence is given by the sandwiched R\'enyi divergence \cite{hiai2017different}.
The detailed proof is deferred to Appendix~\ref{app:REG}.

\medskip
Note that the $n$-shot $\alpha$-leakage $\mathcal{L}_{\alpha}(X^{n} \rightarrow B^{n})_{\rho^{\otimes n}}$ is evaluated under i.i.d.~non-sensitive data $X^n$ as input.
Surprisingly, below we show that when calculating the regularized maximal $\alpha$-leakage $\mathcal{L}_{\alpha}^{\max} (X^n \rightarrow B^n)_{\rho^n}$ for $\alpha >1$, the privacy mechanism $\mathscr{N}_{\mathcal{X}\rightarrow B}^{\otimes n}$ works for general correlated data $X^n$.
Hence, the i.i.d.~assumption on $X^n$ is no longer needed.



Theorem~\ref{thm:regularized_capacity} below establishes the regularized maximal $\alpha$-leakage for $\alpha>1$ and provides an operational meaning for sandwiched R\'enyi capacity $C^*_{\alpha}(\mathscr{N}_{\mathcal{X} \to B})$.

\begin{theorem}[Characterization of regularized maximal $\alpha$-leakage]
\label{thm:regularized_capacity}
For any $\alpha > 1$, the regularized maximal $\alpha$-leakage from $X^n$ with any arbitrary distribution $p_{X^n} \in \mathcal{P}(\mathcal{X}^n)$ that has full support
to $B^n$ under $\mathscr{N}_{\mathcal{X}\rightarrow B}^{\otimes n}$
is given by
\begin{equation}
    \lim_{n\rightarrow\infty} \frac{1}{n} \mathcal{L}_{\alpha}^{\max} (X^n \rightarrow B^n)_{\rho^n} = C^*_{\alpha}(\mathscr{N}_{\mathcal{X} \to B})
    = R^*_{\alpha}(\mathscr{N}_{\mathcal{X} \to B}),
\end{equation}
where $\rho^n_{X^n B^n}:= \sum_{x^n \in \mathcal{X}^n} {p}_{X^n}(x^n) |x^n\rangle\langle x^n| \otimes \rho_{B^n}^{x^n}$.
\end{theorem}

Note that Theorem~\ref{thm:max_alpha_leakage} already characterizes that
\begin{align}
\frac{1}{n} \mathcal{L}_{\alpha}^{\max} (X^n \rightarrow B^n)_{\rho^n}
= 
\frac{1}{n} C_{\alpha}^{\mathrm{A},\mathds{M}}(\mathscr{N}^{\otimes n}_{\mathcal{X} \to B}) 
= \frac{1}{n} C_{\alpha}^{\mathds{M}}(\mathscr{N}^{\otimes n}_{\mathcal{X}\to B}), \quad \forall\, n\in\mathds{N}.
\end{align}
Below we show that the regularized measured R\'enyi capacity is given by the sandwiched R\'enyi capacity
$C^*_{\alpha}(\mathscr{N}_{\mathcal{X} \to B})$.
This concludes Theorem~\ref{thm:regularized_capacity}.

\begin{proposition}[Equivalence of regularized Arimoto capacity, regularized R\'enyi capacity, and sandwiched capacity] \label{prop:regularized_capacity}

Let $\mathscr{N}_{X\rightarrow B}: x\mapsto \rho_{B}^{x}$ denote a classical-quantum channel. For any $\alpha \geq \frac12$, the following identities hold:
\begin{equation} \label{eq:regC}
     \lim_{n\rightarrow\infty}\frac{1}{n} C_{\alpha}^{\mathrm{A},\mathds{M}}(\mathscr{N}^{\otimes n}_{\mathcal{X} \to B}) = \lim_{n\rightarrow\infty}\frac{1}{n} C_{\alpha}^{\mathds{M}}(\mathscr{N}^{\otimes n}_{\mathcal{X}\to B}) = C^*_{\alpha}(\mathscr{N}_{\mathcal{X} \to B})
     = R^*_{\alpha}(\mathscr{N}_{\mathcal{X} \to B}).
\end{equation}
\end{proposition}
\noindent The proof is deferred to Appendix~\ref{app:proof_regularized_capacity}.

\section{Discussion} \label{sec:dis}
In this work, we propose the definition of $\alpha$-leakage and maximal $\alpha$-leakage for a quantum privacy mechanism based on maximal expected $\alpha$-gain. We prove that
(i) one-shot $\alpha$-leakage is determined by measured Arimoto information (Theorem~\ref{thm:alpha_leakage}); 
(ii) one-shot maximal $\alpha$-leakage is determined by measured R\'enyi capacity, measured Arimoto capacity, and measured R\'enyi divergence radius (Theorem~\ref{thm:max_alpha_leakage});
(iii) regularized $\alpha$-leakage is characterized by $\alpha$-tilted sandwiched R\'enyi information (Theorem~\ref{thm:regularizedMI});
(iv) regularized maximal $\alpha$-leakage is characterized by one-shot sandwiched R\'enyi capacity (Theorem~\ref{thm:regularized_capacity}).
Also, we derive various properties (Theorem~\ref{thm:properties}, \ref{thm:sub_Markov}) for maximal $\alpha$-leakage, such as DPI and sub-additivity under the same quantum privacy mechanism. 
The established characterizations apply to the so-called \textit{privacy-utility trade-off} (PUT) scenario of classical data protected by quantum privacy mechanisms, which are depicted as below.

The goal of a privacy mechanism is to preserve information leakage subject to some desired level of utility. By adopting maximal $\alpha$-leakage $\mathcal{L}_\alpha^{\textnormal{max}}(X\to B)_{\rho}$ (Def.~\ref{def:max-alpha-Leakage}) as the privacy metric and any quantum-classical distortion $d(X,B)_\rho$ (see e.g.~\cite{datta2013quantum}) as the utility metric bounded by the maximal permitted distortion $\delta$, we can model this {privacy-utility tradeoff} (PUT) problem as the optimization problem below: for any probability distribution $p_X$ on $\mathcal{X}$,
\begin{subnumcases}
    {}\min_{ x\mapsto \rho_B^x} \hspace{1em} &$\mathcal{L}_\alpha^{\textnormal{max}}(X\to B)_{\rho}$\\
    \textnormal{subject to} \hspace{1em} &$d(X,B)_\rho \leq \delta$,
\end{subnumcases}
the optimal PUT is denoted as 
\begin{equation}
    \mathrm{PUT}(\delta)_\rho := \inf_{d(X,B)_\rho \leq \delta} \mathcal{L}_\alpha^{\textnormal{max}}(X\to B)_{\rho}.
\end{equation}
We have shown that for $\alpha > 1$, the objective function of maximal $\alpha$-leakage $\mathcal{L}_\alpha^{\textnormal{max}}(X\to B)$ is concave in optimization variable $\bar{p}_X$ and quasi-convex in $\rho_B^x$ given $\bar{p}_X$ in Theorem~\ref{thm:properties}. Therefore, if the distortion function $d(X,B)_\rho$ is convex in $\rho_B^x$, this problem belongs to quasi-convex programs \cite{agrawal2020disciplined}. However, given the released quantum state $B$, does our framework differ from measuring it and then tackling it directly with classical techniques? To address this issue, we can consider the Markov chain $X-B-Y$, where $Y$ is a classical state obtained by measuring $B$. Observing that $\mathcal{H}_Y \subseteq \mathcal{H}_B$, we immediately obtain 
\begin{equation}
    \mathrm{PUT}(\delta)_{\rho_{XB}} \leq \mathrm{PUT}(\delta)_{\rho_{XY}}
\end{equation}
due to contraction of the constraint set. Therefore, we suppose that the classical method cannot outperform our quantum privacy mechanism in PUT problem with the privacy metric of maximal $\alpha$-leakage.

Future research directions for this work are abundant, especially for those tasks requiring privacy analysis. Here we list some open problems:

\begin{enumerate}
    \item This work studies the information leakage via a quantum privacy mechanism with a c-c-q Markov chain. The fully-quantum case (q-q-q) or other frameworks (e.g.~c-q-q, q-c-q Markov chain) have not been explore yet.
    
    
    \item Liao \textit{et al.}~\cite{liao2019tunable} proposed the definition of $f$-leakage and maximal $f$-leakage, and formulated them as PUT problems. On the other hand, Hiai and Mosonyi defined measured $f$-divergence \cite{hiai2017different}, one can also consider extending classical $f$-leakage and maximal $f$-leakage to PUT problem under a quantum privacy mechanism by measured $f$-divergence. 
    
    \item 
    For $\alpha = 1$, we find that both $1$-leakage and maximal $1$-leakage are determined by the \textit{measured Arimoto mutual information}
    \begin{align}
    I_1^{\mathrm{A},\mathds{M}}(X;B)_\rho = H_1(X)_p + \inf_{\sigma_B\in \mathcal{S}(\mathcal{H}_B)} D_1^{\mathds{M}}(\rho_{XB}\rVert\mathds{1}_X\otimes\sigma_B)
    \end{align}
    Note that one can also define the \textit{measured mutual information} as
    \begin{align}
    I_1^{\mathds{M}}(X;B)_{\rho}
    = \inf_{\sigma_B\in \mathcal{S}(\mathcal{H}_B)} 
    D_1^{\mathds{M}}\left( \rho_{XB} \rVert \rho_X \otimes \sigma_B \right).
    \end{align}
    In the classical setting, the two quantities coincide. 
    It would be interesting to see if they are identical in the quantum scenario.


    \item In Ref.~\cite{hanson2021guesswork}, it was pointed out that the measured R\'enyi conditional entropy $H_\alpha^{\mathds{M}}(X|B)_{\rho}$ is efficiently computable via the variational expressions \cite{berta2017variational}.
    It would be interesting to find a concrete algorithm for computing this minimax optimization.
\end{enumerate}

\section*{Acknowledgment}
H.-C.~Cheng is supported by the Young Scholar Fellowship (Einstein Program) of the National Science and Technology Council, Taiwan (R.O.C.) under Grants No.~NSTC 112-2636-E-002-009, No.~NSTC 112-2119-M-007-006, No.~NSTC 112-2119-M-001-006, No.~NSTC 112-2124-M-002-003, by the Yushan Young Scholar Program of the Ministry of Education, Taiwan (R.O.C.) under Grants No.~NTU-112V1904-4 and by the research project ``Pioneering Research in Forefront Quantum Computing, Learning and Engineering'' of National Taiwan University under Grant No. NTU-CC-112L893405 and NTU-CC-113L891605. H.-C.~Cheng acknowledges the support from the “Center for Advanced Computing and Imaging in Biomedicine (NTU-113L900702)” through The Featured Areas Research Center Program within the framework of the Higher Education Sprout Project by the Ministry of Education (MOE) in Taiwan.

\begin{appendices}
\section{The Expected \texorpdfstring{$\alpha$}{alpha}-Loss when \texorpdfstring{$\alpha < 1$}{alpha < 1}} \label{app:alpha-loss<1}
\begin{definition}[Expected $\alpha$-loss]
Consider a c-q state $\rho_{XB}:=\sum_x p_X(x)\ket{x}\!\!\bra{x}\otimes\rho_B^x$.
For any $\alpha \in (0,\infty]$, we define an \textit{expected $\alpha$-loss} of POVM $\Pi_{XB}$ under $\rho_{XB}$ as:

$\varepsilon_\alpha (\Pi_{XB})_\rho 
    := $
\begin{subnumcases}
    { }
    \frac{\alpha}{\alpha-1}\left(1 - \Tr\left[ \rho_{XB} \Pi_{XB}^{\frac{\alpha-1}{\alpha}} \right]\right), & if $\alpha > 1$ or $\textnormal{\texttt{supp}}(\rho_{XB}) \subseteq \textnormal{\texttt{supp}}(\Pi_{XB})$\textnormal{;}\label{eq:app-expected-loss}\\
    -\Tr\left[ \rho_{XB} \log\Pi_{XB} \right], & if $\alpha = 1$ and $\textnormal{\texttt{supp}}(\rho_{XB}) \subseteq \textnormal{\texttt{supp}}(\Pi_{XB})$\textnormal{;}\\
    +\infty, & otherwise.
\end{subnumcases}

\end{definition}

\begin{remark}
    Note that expected $\alpha$-loss defined in \eqref{eq:app-expected-loss} is always non-negative. For instance, when $\alpha \in (0,1)$, since $\Pi_{XB}^\beta \geq \mathds{1}_{XB}$ for any $\beta < 0$, one may see that both $\frac{\alpha}{\alpha-1}$ and $1-\Tr[\rho_{XB}\Pi_{XB}^{\frac{\alpha-1}{\alpha}}]$ are negative, producing a positive value of expected $\alpha$-loss.
\end{remark}

\begin{remark}
    For $\alpha = \sfrac12$, $\varepsilon_{\sfrac12}(\Pi_{XB}) = \Tr\left[ \rho_{XB} \Pi_{XB}^{-1} \right] - 1 $ is called \textit{expected exponential-loss} \cite{sypherd2022alphaloss}.
\end{remark}

\begin{remark}
To extend the definition of $\alpha$-loss to $(0,1)$, one may introduce an \emph{$\alpha$-loss operator} of a POVM $\Pi_{XB}$ defined as
 \begin{subnumcases}
    {\ell_\alpha(\Pi_{XB}) := }\frac{\alpha}{\alpha-1}\left(\mathds{1}_{XB} - \Pi_{XB}^{\frac{\alpha-1}{\alpha}}\right), & if $\alpha > 1$ or $\Pi_{XB} > 0$;\\
    -\log \Pi_{XB}, & if $\alpha = 1$ and $\Pi_{XB} > 0$;\\
    +\infty, & otherwise,
\end{subnumcases}
such that $\varepsilon_\alpha(\Pi_{XB})_\rho = \langle \rho_{XB}, \ell_\alpha(\Pi_{XB})\rangle$.
\end{remark}

\section{Proof of Lemma~\ref{lemma:div_radius}} \label{app:div_radius}
\begin{proof}
For $\alpha \geq 0$, the equivalence of measured R\'enyi capacity and measured R\'enyi divergence radius is given by
\begin{align}
    &\hspace{1.5em} C_{\alpha}^{\mathds{M}}(\mathscr{N}_{\mathcal{X} \to B})  = \sup_{p_X \in \mathcal{P}(\mathcal{X})} I_{\alpha}^{\mathds{M}}(X:B)_\rho\\
    &= \sup_{p_X \in \mathcal{P}(\mathcal{X})}\inf_{\sigma_B\in\mathcal{S}(\mathcal{H}_B)} D_\alpha^{\mathds{M}}(\rho_{XB}\rVert\rho_X\otimes\sigma_B)\\
    &= \sup_{p_X \in \mathcal{P}(\mathcal{X})}\inf_{\sigma_B\in\mathcal{S}(\mathcal{H}_B)} \frac{1}{\alpha - 1} \log \mathds{E}_{x\sim p_X}\left[ Q_\alpha^{\mathds{M}}(\rho_B^x\rVert\sigma_B) \right]\\
    &\overset{(\textnormal{a})}{=} \label{eq:minimax} \inf_{\sigma_B\in\mathcal{S}(\mathcal{H}_B)} \sup_{p_X \in \mathcal{P}(\mathcal{X})} \frac{1}{\alpha - 1} \log \mathds{E}_{x\sim p_X}\left[ Q_\alpha^{\mathds{M}}(\rho_B^x\rVert\sigma_B) \right]\\
    &= \label{eq:infor-radius} \inf_{\sigma_B\in \mathcal{S}(\mathcal{H}_B)} \sup_{x\in\mathcal{X}} D_\alpha^{\mathds{M}}(\rho_B^x\rVert\sigma_B) =: R_{\alpha}^{\mathds{M}}(\mathscr{N}_{\mathcal{X} \to B}),
\end{align}
where (a) results from the minimax theorem \cite[Lemma. 2.7]{mosonyi2017strong}, following the $\alpha \in [\frac{1}{2},1)$ case proved in \cite[Lemma~1]{schindler2023continuity}.

Meanwhile, measured Arimoto capacity ($(t)=\mathds{M}$) and sandwiched Arimoto capacity ($(t)=*$) can be expressed as
\begin{align}
    &\hspace{1.5em} C_{\alpha}^{\mathrm{A},{(t)}}(\mathscr{N}_{\mathcal{X} \to B}) = \sup_{p_X \in \mathcal{P}(\mathcal{X})} I_{\alpha}^{\mathrm{A},{(t)}}(X:B)_\rho\\
    &= \sup_{p_X \in \mathcal{P}(\mathcal{X})} D_\alpha^{(t)}(\rho_{XB}\rVert \mathds{1}_X\otimes\sigma_B) + H_\alpha(X)_p\\
    &= \sup_{p_X \in \mathcal{P}(\mathcal{X})} \inf_{\sigma_B} \frac{1}{\alpha-1}\log \sum_{x\in\mathcal{X}}p_X(x)^\alpha Q_\alpha^{(t)}(\rho_B^x\rVert \sigma_B) - \frac{1}{\alpha-1}\log\sum_{x\in\mathcal{X}} p_X(x)^\alpha\\
    &= \sup_{p_X \in \mathcal{P}(\mathcal{X})} \inf_{\sigma_B} \frac{1}{\alpha-1}\log \sum_{x\in\mathcal{X}}\frac{p_X(x)^\alpha}{\sum_{x\in\mathcal{X}}p_X(x)^\alpha} Q_\alpha^{(t)}(\rho_B^x\rVert \sigma_B)\\
    &= \sup_{p_X \in \mathcal{P}(\mathcal{X})} \inf_{\sigma_B} \frac{1}{\alpha-1}\log \sum_{x\in\mathcal{X}} p_X^{(\alpha)}(x) Q_\alpha^{(t)}(\rho_B^x\rVert \sigma_B) \\
    &\overset{(\textnormal{b})}{=} \sup_{p_X \in \mathcal{P}(\mathcal{X})} \inf_{\sigma_B} \frac{1}{\alpha-1}\log \sum_{x\in\mathcal{X}} p_X(x) Q_\alpha^{(t)}(\rho_B^x\rVert \sigma_B) \label{eq:lemma6_conti}\\
    &= \sup_{p_{X} \in \mathcal{P}(\mathcal{X})} I_\alpha^{(t)}(X:B)_\rho =: C_\alpha^{(t)}(\mathscr{N}_{\mathcal{X} \to B}), \label{eq:thm2_upR}
\end{align}
where (b) comes from Lemma~\ref{lemma:alpha-tilted} since the objective function in \eqref{eq:lemma6_conti} is continuous (see e.g. \cite[Lemma 1]{CGH18}) in $p_X$.

Thus, we show the equivalence not only between  measured R\'enyi capacity, measured R\'enyi divergence radius and measured Arimoto capacity for $\alpha \geq 0$, but also the equivalence between sandwiched R\'enyi capacity, sandwiched R\'enyi divergence radius and sandwiched Arimoto capacity for $\alpha \geq \sfrac{1}{2}$.
\end{proof}

\section{Proof of Theorem~\ref{theorem:alpha-gain}} \label{app:alpha_gain}
\begin{proof} 
We prove Theorem~\ref{theorem:alpha-gain} via Lagrange duality and the variational expression in Lemma~\ref{lemma:var_measure}.
Note that we only consider Definitions~\ref{def:alpha-gain} and \ref{def:max-alpha-gain} for $\alpha \geq 1$ (because later we only focus on the $\alpha$-leakage in Definition~\ref{def:alpha-Leakage} for such a range), we remark that Theorem~\ref{theorem:alpha-gain} holds for $\alpha \in [\sfrac12, \infty]$.
In the following, we first prove the $\alpha \in [\sfrac12, \infty] \backslash \{1\}$ case. 
Recalling Definition~\ref{def:max-alpha-gain} of the maximal expected $\alpha$-gain, we first modify the optimization region:
\begin{align}
    \frac{\alpha}{\alpha-1}\log\mathsf{P}_\alpha(X|B)_\rho 
    &\overset{\textnormal{(a)}}{=} \label{eq:thm1_equality_constraint} \sup_{\substack{\omega_{XB} > 0\\ \Tr_X[\omega_{XB}] = \mathds{1}_B}} \frac{\alpha}{\alpha-1}\log  \Tr \left[ \rho_{XB} \omega_{XB} ^{\frac{\alpha-1}{\alpha}}\right]
    \\
    &\overset{\textnormal{(b)}}{=} \label{eq:thm1_part1} \sup_{\substack{\omega_{XB} > 0\\ \Tr_X[\omega_{XB}] \leq \mathds{1}_B}} \frac{\alpha}{\alpha-1}\log  \Tr \left[ \rho_{XB} \omega_{XB} ^{\frac{\alpha-1}{\alpha}}\right],
\end{align}
where (a) is because the objective function is continuous and hence we can take the interior of the finite-dimensional POVMs $\omega_{XB}$;
(b) follows from the fact that $(\cdot)^\frac{\alpha-1}{\alpha}$ is operator monotone increasing for $\alpha > 1$ and operator monotone decreasing for $\alpha \in [\sfrac{1}{2},1)$ \cite[\S 4]{Hiai2014matrix}.

We introduce Lagrange operator $W_B \geq 0$ to rewrite \eqref{eq:thm1_part1} into its dual problem:
\begin{align}
    &\quad \label{eq:thm1_part2} \inf_{W_B \geq 0} \sup_{\omega_{XB} > 0} \frac{\alpha}{\alpha-1}\log\Tr \left[ \rho_{XB} \omega_{XB} ^{\frac{\alpha-1}{\alpha}} \right] -  \left(\Tr[W_B\Tr_X[\omega_{XB}]]-\Tr[W_B]\right).
\end{align}
Note that strong duality theorem \cite[Prop.~5.3.1]{bertsekas1995nonlinear} holds since there exists at least an interior point (e.g.~$\omega_{XB} = \frac{\mathds{1}_X}{2|\mathcal{X}|} \otimes \mathds{1}_B$) satisfying the constraint $\Tr_X[\omega_{XB}] \leq \mathds{1}_B$.

Next, we rewrite the penalty terms in  \eqref{eq:thm1_part2} as follows
\begin{align}
    &\quad \label{eq:thm1_0log} \inf_{W_B \geq 0} \sup_{\omega_{XB} > 0} \frac{\alpha}{\alpha-1}\log\Tr \left[ \rho_{XB} \omega_{XB} ^{\frac{\alpha-1}{\alpha}} \right] + \big((1-\Tr[W_B\Tr_X[\omega_{XB}]]) - (1-\Tr[W_B])\big)
    \\
    &\overset{\textnormal{(c)}}{=} \label{eq:thm1_1log} \inf_{W_B \geq 0} \sup_{\omega_{XB} > 0} \frac{\alpha}{\alpha-1}\log\Tr \left[ \rho_{XB} \omega_{XB} ^{\frac{\alpha-1}{\alpha}} \right] + \big((-\log\Tr[W_B\Tr_X[\omega_{XB}]]) - (1-\Tr[W_B])\big)\\
    &\overset{\textnormal{(d)}}{=} \label{eq:thm1_part3} \inf_{W_B \geq 0} \sup_{\omega_{XB} > 0} \frac{\alpha}{\alpha-1}\log\Tr \left[ \rho_{XB} \omega_{XB} ^{\frac{\alpha-1}{\alpha}} \right] + \big((-\log\Tr[W_B\Tr_X[\omega_{XB}]]) - (-\log\Tr[W_B])\big).
\end{align}
Here we first show that the penalty $1-\Tr[W_B\Tr_X[\omega_{XB}]]$ can be changed to $-\log\Tr[W_B\Tr_X[\omega_{XB}]]$ in identity (c).
Observe that \eqref{eq:thm1_1log} is invariant under the substitution $\omega_{XB} \mapsto \gamma \omega_{XB}$ for $\gamma > 0$. 
Further, note that the optimizer $W_B^\star$ never attains value at zero in \eqref{eq:thm1_1log}; otherwise, the optimization value of expression \eqref{eq:thm1_1log} escalates to $+\infty$.
Because $\Tr[W_B\Tr_X[\omega_{XB}]] > 0$ for $W_B>0$ and $\omega_{XB}>0$,
we can impose a normalization constraint $\Tr[W_B\Tr_X[\omega_{XB}]] = 1$ on the following equation:
\begin{align}
    &\quad \inf_{W_B \geq 0} \sup_{\omega_{XB} > 0} \frac{\alpha}{\alpha-1}\log\Tr \left[ \rho_{XB} \omega_{XB} ^{\frac{\alpha-1}{\alpha}} \right] + \big((-\log\Tr[W_B\Tr_X[\omega_{XB}]]) - (1-\Tr[W_B])\big)\\
    &= \inf_{W_B \geq 0} \sup_{\substack{\omega_{XB} > 0\\\Tr[W_B\Tr_X[\omega_{XB}] = 1}} \frac{\alpha}{\alpha-1}\log\Tr \left[ \rho_{XB} \omega_{XB} ^{\frac{\alpha-1}{\alpha}} \right] + \big((-\log\Tr[W_B\Tr_X[\omega_{XB}]]) - (1-\Tr[W_B])\big)\\
    &\leq \inf_{W_B \geq 0} \sup_{\omega_{XB} > 0} \frac{\alpha}{\alpha-1}\log\Tr \left[ \rho_{XB} \omega_{XB} ^{\frac{\alpha-1}{\alpha}} \right] + \big((1-\Tr[W_B\Tr_X[\omega_{XB}]]) - (1-\Tr[W_B])\big),
\end{align}
where the inequality comes from relaxation of the constraint set and $-\log x = 1 - x$ when $x = 1$.
On the other hand, since we have $\log(x + 1) \leq x$ for all $x > -1$, $-\log\Tr[W_B\Tr_X[\omega_{XB}] \geq 1 - \Tr[W_B\Tr_X[\omega_{XB}]$, we can also lower-bound \eqref{eq:thm1_1log} by \eqref{eq:thm1_0log}.
Thus, we have proved the identity (c).
Now we can apply the same technique to further prove identity (d) by observing that \eqref{eq:thm1_part3} is invariant under the substitution $W_B \mapsto \zeta W_B$ for any $\zeta > 0$.
We remark that a similar technique was used in the proof of \cite[Lemma 1]{berta2017variational}.

Finally, let $\sigma_B = \frac{W_B}{\Tr[W_B]}$ in \eqref{eq:thm1_part3}, we obtain
\begin{align}
    &\quad \inf_{\sigma_B \in \mathcal{S}(\mathcal{H}_B)} \sup_{\omega_{XB} > 0} \frac{\alpha}{\alpha-1}\log\Tr \left[ \rho_{XB} \omega_{XB} ^{\frac{\alpha-1}{\alpha}} \right] -\log\Tr[\sigma_B\Tr_X[\omega_{XB}]]\\
    &= \inf_{\sigma_B \in \mathcal{S}(\mathcal{H}_B)} \sup_{\omega_{XB} > 0} \frac{\alpha}{\alpha-1}\log\Tr \left[ \rho_{XB} \omega_{XB} ^{\frac{\alpha-1}{\alpha}} \right] -\log\Tr[(\mathds{1}_X \otimes \sigma_B)\omega_{XB}]\\
    &\overset{\textnormal{(e)}}{=} \label{eq:thm1_part4} \inf_{\sigma_B \in \mathcal{S}(\mathcal{H}_B)} D_\alpha^{\mathds{M}}(\rho_{XB} \rVert \mathds{1}_X \otimes \sigma_B)\\
    &= -H_\alpha^{\mathds{M}}(X|B)_\rho,
\end{align}
where (e) is from the variational formula of measured R\'enyi divergence (Lemma~\ref{lemma:var_measure}).

\end{proof}

\section{Proof of Theorem~\ref{thm:alpha_leakage}} \label{app:alpha_leakage}
\begin{proof}
For $\alpha \in (1,\infty]$, by definition, the left-hand side of \eqref{eq:thm 1} can be written as
\begin{align} \label{eq:split-alpha}
    \mathcal{L}_{\alpha} (X \rightarrow B)_\rho &= \frac{\alpha}{\alpha-1} \log \mathsf{P}_\alpha(X|B)_\rho -
    \frac{\alpha}{\alpha-1} \log \sup_{p_{\hat{X}}} \sum_{x\in\mathcal{X}}  p_X(x) {p_{\hat{X}}}(x)^{\frac{\alpha-1}{\alpha}}\\
    &\overset{(\textnormal{a})}{=} \left(-H_\alpha^\mathds{M}(X|B)_\rho\right) -\left(-H_\alpha(X)_p\right) = I_{\alpha}^{\mathrm{A},\mathds{M}}(X:B)_\rho,
\end{align}
where (a) follows from Theorem~\ref{theorem:alpha-gain} indicates that the first term of \eqref{eq:split-alpha} can be expressed as the measured conditional R\'enyi entropy, whereas the latter term in \eqref{eq:split-alpha} can be simplified by the KKT condition \cite[Chapter 5.5.3]{boyd2004convex}.

On the other hand, the $\alpha = 1$ case can be proved as follows.
\begin{align}
    &\hspace{1.5em} \mathcal{L}_1(X\to B)_\rho = -\varepsilon_1(X|B)_\rho + \varepsilon_1(X)_\rho\\
    &\overset{(\textnormal{b})} {=} \label{eq:thm2_log-loss} \left(-H_1^\mathds{M}(X|B)_\rho\right) -\left(-H_1(X)_p\right) = I_1^{\mathrm{A},\mathds{M}}(X:B)_\rho,
\end{align}
where (b) follows from the similar Lagrange multiplier method as the $\alpha > 1$ proof in Appendix~\ref{app:alpha_gain}:
\begin{align} \label{eq:thm2_neg-logloss}
    &\hspace{1.5em} \varepsilon_1(X|B)_\rho =  \inf_{\textnormal{POVM }\Pi_{XB}}-\Tr[\rho_{XB}\log\Pi_{XB}]\\
    &= -\inf_{\sigma_B\in\mathcal{S}(\mathcal{H}_B)} \sup_{\Pi_{XB} > 0} \left( \Tr\left[\rho_{XB} \log\Pi_{XB} \right] - \log\Tr[(\mathds{1}_X\otimes\sigma_B)\Pi_{XB}] \right)\\
    &= H_1^{\mathds{M}}(X|B)_\rho.
\end{align}

\end{proof}

\section{Proof of Theorem~\ref{thm:max_alpha_leakage}} \label{app:max_alpha_leakage}
\begin{proof}
For the $\alpha = 1 $ case, by Definition~\ref{def:max-alpha-Leakage} and Theorem~\ref{thm:alpha_leakage}, we have
\begin{equation}
     \mathcal{L}_{1}^{\max} (X \to B)_\rho = \sup_{p_{S|X}: S-X-B} I_1^{\mathrm{A},\mathds{M}}(S:B)_\rho\leq I_1^{\mathrm{A},\mathds{M}}(X:B)_\rho,\label{eq:alpha=1,DPI}
\end{equation}
where the inequality is from Lemma~\ref{lemma:DPI_measure}. To prove the achievability of \eqref{eq:alpha=1,DPI}, we set $\rho_{SB}= \rho_{XB}$, and thus $\mathcal{L}_{1}^{\max} (X \to B)_\rho = I_1^{\mathrm{A},\mathds{M}}(X:B)_\rho.$ 

Now we consider the scenario of $\alpha > 1$. This proof consists of two parts: the upper bound is optimality, whereas the lower bound is achievability.

\vspace{2em}

\noindent\textbf{Upper bound} (optimality):
Subsequently, we fix the distribution $p_X$ and the collection of states $\{\rho_B^x\}_{x\in\mathcal{X}}$.
We first rewrite the maximal $\alpha$-leakage as the following:
\begin{align}
    \mathcal{L}_\alpha^{\max} (X \to B)_\rho &= \sup_{p_{S|X}: S-X-B} I_{\alpha}^{\mathrm{A},\mathds{M}}(S:B)_\rho \label{eq:thm3_nobar}\\
    &\overset{}{=} \sup_{\substack{p_{\bar{S}\bar{X}}: p_{\bar{X}} = p_X\\
    \bar{S}-\bar{X}-B }} I_{\alpha}^{\mathrm{A},\mathds{M}}(\bar{S}:B)_\rho, \label{eq:thm3_bar}
\end{align}
where the supremum in \eqref{eq:thm3_nobar} is over the set of Markov classical-classical-quantum states $\rho_{SXB}$ of the form:
\begin{align} \label{eq:set_thm3_optnobar}
\left\{ \rho_{SXB} = \sum_s \sum_{ x\in\mathcal{X}}p_{S|X=x}(s|x)|s\rangle \langle s| \otimes  p_{X}(x) |x\rangle \langle x| \otimes \rho_B^x 
: \forall\, \{ p_{S|X=x} \}_{x\in\mathcal{X}} \subseteq \mathcal{P(S)}
\right\},
\end{align}
whereas the supremum in \eqref{eq:thm3_bar} is over the set of states $\rho_{\bar{S}\bar{X}B}$ of the form:
\begin{align} \label{eq:set_thm3_optbar}
\Bigg\{
\rho_{\bar{S}\bar{X}B} = \sum_{{s} }\sum_{ {x} \in \bar{\mathcal{X}}} p_{\bar{S}}({s})\ket{{s}}\!\!\bra{{s}} \otimes p_{\bar{X}|\bar{S}}({x}|{s}) \ket{{x}}\!\!\bra{{x}} \otimes \rho_B^{{x}} :
\forall\ p_{\bar{S}} \in \mathcal{P}(\bar{\mathcal{S}}),
p_{\bar{X}|\bar{S}} \text{~s.t.~} p_{\bar{X}} = p_X \Bigg\}.
\end{align}
By inspection, the sets \eqref{eq:set_thm3_optnobar} and \eqref{eq:set_thm3_optbar} coincide.

Next, we relax the constraint $p_{\bar{X}}  = p_X $ for the stochastic transformation $p_{\bar{X}|\bar{S}}$ in \eqref{eq:set_thm3_optbar}  to all distributions on the support of $p_X$, i.e.
\begin{align}
    \mathcal{L}_\alpha^{\max} (X \to B)_\rho 
    &\leq\sup_{\substack{
    p_{\bar{X}|\bar{S}}(\cdot|s)\in\mathcal{P}( \textnormal{\texttt{supp}}(p_X))}} \sup_{p_{\bar{S}}: \bar{S}-\bar{X}-B} I_{\alpha}^{\mathrm{A},\mathds{M}}(\bar{S}:B)_\rho \label{eq:thm3_2sup}\\ 
    &\overset{(\textnormal{a})}{=} \sup_{\substack{
    p_{\bar{X}|\bar{S}}(\cdot|s)\in\mathcal{P}( \textnormal{\texttt{supp}}(p_X))}} \sup_{p_{\bar{S}}: \bar{S}-\bar{X}-B} I_{\alpha}^{\mathds{M}}(\bar{S}:B)_\rho\label{eq:thm2_upSlast}\\
    &\overset{(\textnormal{b})}{=} \sup_{p_{\bar{X}} \in \mathcal{P}(\textnormal{\texttt{supp}}(p_X))} I_{\alpha}^{\mathds{M}}(\bar{X}:B)_\rho = C_{\alpha}^{\mathds{M}}(\mathscr{N}_{\textnormal{\texttt{supp}}(p_X) \to B}) \\
    &\overset{(\textnormal{c})}{=} \sup_{p_{\bar{X}} \in \mathcal{P}(\textnormal{\texttt{supp}}(p_X))} I_{\alpha}^{\mathrm{A},\mathds{M}}(\bar{X}:B)_\rho = C_{\alpha}^{\mathrm{A},\mathds{M}}(\mathscr{N}_{\textnormal{\texttt{supp}}(p_X) \to B})\\
    &\overset{(\textnormal{d})}{=}R_{\alpha}^{\mathds{M}}(\mathscr{N}_{\textnormal{\texttt{supp}}(p_X) \to B}) \label{eq:thm2_uplast}, 
\end{align}
where (a) holds from that the supremum of measured Arimoto information equals the supremum measured R\'enyi information (Lemma~\ref{lemma:div_radius});
(b) results from the data processing inequality
(Lemma~\ref{lemma:DPI_measure}) of measured R\'enyi information for the Markov chain $\bar{S}-\bar{X}-B$, where the equality is attained when $\bar{S} = \bar{X}$.\footnote{Indeed the data-processing operation of Lemma~\ref{lemma:DPI_measure} is expressed in terms of a channel.
For the Markov chain  $\bar{S}-\bar{X}-B$ of the form 
in \eqref{eq:set_thm3_optbar}, we can regard the stochastic transformation $p_{\bar{S}|\bar{X}}$ as a channel $\mathcal{N}_{\textnormal{\texttt{supp}}(p_X)\to \bar{S}}
=
\sum_s \sum_{x\in\textnormal{\texttt{supp}}(p_X)} p_{\bar{S}|X}(s|x) |s\rangle \langle x| \cdot |x\rangle \langle s|
$.
Hence, Lemma~\ref{lemma:DPI_measure} implies the following inequality $I_\alpha^{\mathds{M}}(\bar{S}:B)_\rho = \inf_{\sigma_B} D_\alpha^{\mathds{M}}(\rho_{\bar{S}B} \rVert \rho_{\bar{S}} \otimes \sigma_B)
= \inf_{\sigma_B} D_\alpha^{\mathds{M}}\left((\mathcal{N}_{\textnormal{\texttt{supp}}(p_X)\to \bar{S}} \otimes \text{id}_B)(\rho_{\bar{X}B}) \rVert(\mathcal{N}_{\textnormal{\texttt{supp}}(p_X)\to \bar{S}} \otimes \text{id}_B) (\rho_{\bar{X}} \otimes \sigma_B) \right)
\leq \inf_{\sigma_B} D_\alpha^{\mathds{M}}(\rho_{\bar{X}B} \rVert \rho_{\bar{X}} \otimes \sigma_B) = I_\alpha^{\mathds{M}}(\bar{X}:B)_\rho$. 
}
The last two lines, (c) and (d), are again direct consequences of Lemma~\ref{lemma:div_radius}.

\vspace{1.5em}

\noindent\textbf{Lower bound} (achievability): 
 We bound the maximal $\alpha$-leakage from below by considering a random variable $S$, where $S-X-B$ forms a classical-classical-quantum Markov chain and $H_1(X|S) = 0$. Let us apply the similar method as in \cite{liao2019tunable}: let $\mathcal{S}$ and $\mathcal{X}$ be a bijection so that $\mathcal{S}$ is composed of $\mathcal{S}_{x}$ (i.e. $\mathcal{S}= \bigcup_{x \in \mathcal{X} }\mathcal{S}_x$), with $S = s \in \mathcal{S}_x$ if and only if $X=x$. This way, we can construct a bijective function $f: x\mapsto \mathbb{R}$ by $S \sim P_S$ for a random variable $X$ distributed over support $\mathcal{X}$ as 
 \begin{equation}\label{eq:f(x)}
 f(x)=\sum_{s \in \mathcal{S}_x} P_S(s)^{\alpha}.
 \end{equation}
 Then, a probability distribution $p_{\bar{X}}(x)$ over support $\mathcal{X}$ can be formed by \eqref{eq:f(x)} as
 \begin{equation}
     p_{\bar{X}}(x)=\frac{f(x)}{\sum_{x\in\mathcal{X}} f(x)}
     =\frac{\sum_{s \in \mathcal{S}_x}P_S(s)^{\alpha}}{\sum_{x\in\mathcal{X}}\sum_{s \in \mathcal{S}_x} P_S(s)^{\alpha}} \hspace{2em}\forall x \in\mathcal{X}.\label{eq:PbarxPu}
 \end{equation}
And $\rho^{s}_{B}$ is constructed as
\begin{subnumcases}
{\rho^{s}_{B}=}
\rho^{x}_{B}, & for $s \in \mathcal{S}_{x}$\\
0, & otherwise.
\end{subnumcases}

Setting $\sigma_{SB} \equiv \sum_{s\in\mathcal{S}}\frac{1}{\left|\mathcal{S}\right|}\ket{s}\!\!\bra{s}\otimes\sigma_B$ , we have
\begin{align}
&\hspace{1.5em}I_{\alpha}^{\mathds{M}}(\bar{X}:B)_\rho =
\inf_{\sigma_{B}\in\mathcal{S}(\mathcal{H}_{B})} D_{\alpha}^{\mathds{M}}(\rho_{\bar{X}B}\rVert\rho_{\bar{X}}\otimes\sigma_B)\\
&=\inf_{\sigma_{B}\in\mathcal{S}(\mathcal{H}_{B})} \frac{1}{\alpha-1}\log \mathds{E}_{x\sim p_{\bar{X}}}\left[ Q_\alpha^{\mathds{M}}(\rho_B^x\rVert\sigma_{B}) \right]\\
&=\inf_{\sigma_{B}\in\mathcal{S}(\mathcal{H}_{B})} \frac{1}{\alpha-1}\log \sum_{x \in \mathcal{X}} \frac{\sum_{s \in \mathcal{S}_{x}}P_{S}(s)^{\alpha}}{\sum_{x\in\mathcal{X}}\sum_{s \in \mathcal{S}_{x}} P_{S}(s)^{\alpha}}
Q_\alpha^{\mathds{M}}(\rho_B^x\rVert\sigma_{B})\\
&=\inf_{\sigma_{B}\in\mathcal{S}(\mathcal{H}_{B})} \frac{1}{\alpha -1}\log\sum_{s \in \mathcal{S}} P_{S}(s)^{\alpha} Q_\alpha^{\mathds{M}}(\rho_B^s\rVert\sigma_{B})+H_{\alpha}(S)_p\\
&=\inf_{\sigma_{B}\in\mathcal{S}(\mathcal{H}_{B})} \frac{1}{\alpha -1}\log \bigg[ \sup_{\{\Pi^{s}_{B}\}_{s}}\sum_{s\in \mathcal{S}} \left(P_{S}(s)\Tr \left[ \rho_{B}^{s} (\Pi^{s}_{B})^{\frac{\alpha-1}{\alpha}}\right]\right)^{\alpha}\left( \frac{1}{|\mathcal{S}|}\Tr \left[ \sigma_{B} \Pi^{s}_{B}\right]\right)^{1-\alpha} \bigg] \label{eq:supforalpha>1}
\notag\\
&\quad -\log|\mathcal{S}|+H_{\alpha}(S)_p\\
&=\inf_{\sigma_{B}\in\mathcal{S}(\mathcal{H}_{B})} \frac{1}{\alpha -1}\log Q_\alpha^{\mathds{M}}(\rho_{SB}\rVert\sigma_{SB})-\log|\mathcal{S}|+H_{\alpha}(S)_p\\
&=I_{\alpha}^{\mathrm{A},\mathds{M}}(S:B)_\rho.
\end{align}
Therefore,
\begin{align}
    &\hspace{1.5em}\mathcal{L}_\alpha^{\max} (X \to B)_\rho\\
    &= \sup_{p_{S|X}: S-X-B} I_{\alpha}^{\mathrm{A},\mathds{M}}(S:B)_\rho\\  
    &\geq \sup_{\substack{p_{S|X}: S-X-B\\ H_1(X|S)=0}} I_{\alpha}^{\mathrm{A},\mathds{M}}(S:B)_\rho\label{eq:H_1(S|X)=0}\\
    &= \sup_{p_{\bar{X}} \in \mathcal{P}(\textnormal{\texttt{supp}}(p_X))}I_{\alpha}^{\mathds{M}}(\bar{X}:B)_\rho \equiv C_{\alpha}^{\mathds{M}}(\mathscr{N}_{\textnormal{\texttt{supp}}(p_X) \to B}). \label{eq:P_xbar}
\end{align}
The last equality holds since for any $p_{\bar{X}} \in\mathcal{P}( \textnormal{\texttt{supp}}(p_X))$, we can construct corresponding $P_{S}(s)$ for $s\in\mathcal{S}$ that satisfies \eqref{eq:f(x)} and \eqref{eq:PbarxPu} by bijection.
Thus the supremum over $S-X-B$ and $H_1(X|S)=0$ in \eqref{eq:H_1(S|X)=0} is equivalent to the supremum over $p_{\bar{X}}$ in \eqref{eq:P_xbar}.
By combining \eqref{eq:thm2_uplast} and \eqref{eq:P_xbar}, we proved Theorem~\ref{thm:max_alpha_leakage}. We remark that a similar technique was used in the proof of \cite[Thm.~2]{liao2019tunable}.
\end{proof}

\section{Proof of Theorem~\ref{thm:properties}} \label{app:properties}

\begin{proof}
\hspace{1em}\vspace{1em}

\noindent\textbf{The proof of part 1}: For $\alpha > 1$, the maximal $\alpha$-leakage can be expressed as
\begin{equation} \label{eq:thm4-1-obj}
     \mathcal{L}_{\alpha}^{\max} (X \to B)_\rho = \sup_{\bar{p}_X \in \mathcal{P}( \textnormal{\texttt{supp}}(p_X))} \inf_{\sigma_B\in\mathcal{S}(\mathcal{H}_B)} \frac{1}{\alpha - 1} \log \left( \sum_{x\in\mathcal{X}} \bar{p}_X(x) Q_{\alpha}^{\mathds{M}}(\rho_B^x\rVert\sigma_B) \right),
\end{equation}
where the objective function in \eqref{eq:thm4-1-obj} is concave in the optimization variable $\bar{p}_X$.

The proof for $\alpha = 1$ follows from the concavity of measured Arimoto information (Def.~\ref{def:measured-ArimotoMI})
\begin{equation}
I_{\alpha}^{\mathrm{A},\mathds{M}}(X:B)_\rho = H_\alpha(X)_p + \inf_{\sigma_B
\in\mathcal{S}(\mathcal{H}_B)} D^{\mathds{M}}_1(\rho_{XB}\rVert\mathds{1}_X\otimes\sigma_B).
\end{equation}
 
When $\alpha=1$, it is straightforward that Shannon entropy $H_1(X)$ is concave function of $p_X$. 
The concavity of $\inf_{\sigma_B
\in\mathcal{S}(\mathcal{H}_B)} D^{\mathds{M}}_1(\rho_{XB}\rVert\mathds{1}_X\otimes\sigma_B)$ is from that the variational expression (Lemma~\ref{lemma:var_measure}) is linear in $p_X$: 
\begin{equation} \label{eq:thm4-1-concave}
D^{\mathds{M}}_1(\rho_{XB} \rVert \mathds{1}_X\otimes\sigma_B) = \sup_{\omega_{XB} > 0} \Tr \left[ \rho_{XB} \log \omega_{XB} \right] + 1 - \Tr\left[ (\mathds{1}_X\otimes\sigma_B) \omega_{XB}\right].
\end{equation}
Besides, the infimum of arbitrary concave (linear in \eqref{eq:thm4-1-concave}) functions is still concave.
Therefore, the maximal 1-leakage is a concave function of $p_X$.\\

\noindent\textbf{The proof of part 2}:
For two c-q states $\rho_{XB}$ and $\tau_{XB}$ which follow the same distribution $p_X$ and $\alpha \in [1,\infty]$, we assume $D^\mathds{M}_{\alpha} (\rho_{XB}\parallel \rho_X\otimes\sigma_B)\geq D^\mathds{M}_{\alpha} (\tau_{XB}\parallel \rho_X\otimes\sigma_B)$ without loss of generality.
Here we prove the $\alpha = 1$ case.
For a c-q state $\beta \rho_{XB}+(1-\beta)\tau_{XB}$ (with $0 \leq \beta \leq 1$) by Definition~\ref{def:big_def} and Lemma~\ref{lemma:var_measure}, we observe that
\begin{align}
&\hspace{1.1em}\inf_{\sigma_B \in \mathcal{S}(\mathcal{H}_B)} D_1^{\mathds{M}}(\beta \rho_{XB}+(1-\beta)\tau_{XB}\rVert\rho_X\otimes\sigma_B)\\&=\inf_{\sigma_B \in \mathcal{S}(\mathcal{H}_B)} \sup_{\Pi_{XB}>0} \Tr \left[ (\beta \rho_{XB}+(1-\beta)\tau_{XB})\log \Pi_{XB} \right]-\log \Tr\left[(\rho_X\otimes\sigma_B)\Pi_{XB}\right]\\
&\leq \inf_{\sigma_B \in \mathcal{S}(\mathcal{H}_B)} \sup_{\Pi_{XB}>0} \Tr \left[ (\beta \rho_{XB}+(1-\beta)\rho_{XB})\log \Pi_{XB} \right]-\log \Tr\left[(\rho_X\otimes\sigma_B)\Pi_{XB}\right]\\
&=D_1^\mathds{M} (\rho_{XB}\parallel \rho_X\otimes\sigma_B),
\end{align}\\
where the inequality follows from our assumption. Note that the quasi-convexity is preserved under taking infimum \cite{boyd2004convex}.
Thus, combine this with Theorem~\ref{thm:max_alpha_leakage}, maximal $1$-leakage is quasi-convex in $\rho_B^x$ given $\textit{p}_X$.
The proof of maximal $\alpha$-leakage for $\alpha > 1$ follows similarly by using Lemma~\ref{lemma:var_measure} and Theorem~\ref{thm:max_alpha_leakage}.
\vspace{1em} 

\noindent\textbf{The proof of part 3}:
Since the first-order derivative of $\alpha$ in the measured R\'enyi information is non-negative, we immediately see that maximal $\alpha$-leakage is non-decreasing in $\alpha$.
\vspace{1em}

\noindent\textbf{The proof of part 4}:
We can view Markov chain $S-X-B$ as two channels from $S\rightarrow X$ and $X\rightarrow B$, then by Lemma~\ref{lemma:DPI_measure} and Theorem~\ref{thm:max_alpha_leakage}, maximal-$\alpha$ leakage satisfies data processing inequality.

\noindent\textbf{The proof of part 5}:
The lower bound directly follows from the non-negativity of measured R\'enyi divergence.

To derive the upper bound for $\alpha > 1$, we bound maximal $\alpha$-leakage by maximal $\infty$-leakage using monotonicity with respect to tunable parameter $\alpha$ first (Theorem~\ref{thm:properties}, part 3), and adopt the divergence radius expression of maximal $\infty$-leakage to obtain an optimization problem of maximum relative entropy (defined in Eq.~\eqref{eq:max-relative-entropy}):
\begin{align}
    &\quad \mathcal{L}_\alpha^{\max} (X \to B)_\rho \leq \mathcal{L}_\infty^{\max} (X \to B)_\rho \\
    &= R_\infty^{\mathds{M}}(\mathscr{N}_{\textnormal{\texttt{supp}}(p_X)\to B})\\
    &= \log\sup_{x \in \textnormal{\texttt{supp}}(p_X)}\inf_{\sigma_B \in \mathcal{S}(\mathcal{H}_B)}\inf\{\Tr[M]: \Tr[\rho_B^x] \leq \Tr[M\sigma_B]\}\\
    &\leq \log\sum_{x \in \textnormal{\texttt{supp}}(p_X)}\inf_{\sigma_B \in \mathcal{S}(\mathcal{H}_B)}\inf\{\Tr[M]: \Tr[\rho_B^x] \leq \Tr[M\sigma_B]\}\\
    &= \log(|\textnormal{\texttt{supp}}(p_X)|),
\end{align}
where $M$ is a Hermitian operator.
On the other hand, we can bound $\mathcal{L}_1^{\max} (X \to B)_\rho$ by
\begin{align}
    \mathcal{L}_1^{\max} (X \to B)_\rho &= H_1(X)_p - H^{\mathds{M}}_1(X|B)_\rho\\
    &= H_1(X)_p + \inf_{\sigma_B} D^{\mathds{M}}_1(\rho_{XB} \rVert \mathds{1}_X \otimes \sigma_B)\\
    &\leq H_1(X)_p + D^{\mathds{M}}_1(\rho_{XB} \rVert \mathds{1}_X \otimes \rho_B)\\
    &\overset{(*)}{\leq} H_1(X)_p + D^{\mathds{M}}_1(\rho_{XB} \rVert \rho_{XB}) = H_1(X)_p,
\end{align}
where ($*$) follows from the reduction criterion of separability \cite[Sec.~III]{Horodeki1999reduction}.

\end{proof}

\section{Proof of Theorem~\ref{thm:sub_Markov}} \label{app:sub_Markov}
\begin{proof}
Throughout the proof, 
    we denote by 
    \begin{align}
    {\bar{\rho}}_{XB} = \sum_{x \in \mathcal{X}} \bar{p}_{X} |x\rangle \langle x| \otimes \rho_{B}^{x}
    \end{align}
    for any input distribution $\bar{p}_{X} \in \mathcal{P}(\mathcal{X})$ as a dummy variable in optimization.
\begin{align}
    &\hspace{1.5em} \mathcal{L}_\alpha ^{\max} (X \to B_1,B_2)_{\bar{\rho}} = \sup_{{\bar{p}}_X \in \mathcal{P}(\textnormal{\texttt{supp}}(p_X))}I_{\alpha}^{\mathds{M}}(X:B_1  B_2)_\rho\\
    &= \sup_{{\bar{p}}_X \in \mathcal{P}(\textnormal{\texttt{supp}}(p_X))} \inf_{\sigma_{B_1}, \sigma_{B_2}} D_{\alpha}^{\mathds{M}}({\bar{\rho}}_{XB_1B_2} \rVert {\bar{\rho}}_{X} \otimes \sigma_{B_1} \otimes \sigma_{B_2})\\
    &\overset{(\textnormal{a})}{=} \sup_{{\bar{p}}_X \in \mathcal{P}(\textnormal{\texttt{supp}}(p_X))} \left[ \inf_{\sigma_{B_1}} D_{\alpha}^{\mathds{M}}({\bar{\rho}}_{XB_1} \rVert {\bar{\rho}}_{X} \otimes \sigma_{B_1}) + \inf_{\sigma_{B_2}} D_{\alpha}^{\mathds{M}}(\rho_{B_2} \rVert \sigma_{B_2}) \right]\\
    &\overset{(\textnormal{b})}{\leq} \sup_{{\bar{p}}_{X_1} \in\mathcal{P}( \textnormal{\texttt{supp}}(p_X))}  \inf_{\sigma_{B_1}} D_{\alpha}^{\mathds{M}}({\bar{\rho}}_{X_1B_1} \rVert {\bar{\rho}}_{X_1} \otimes \sigma_{B_1}) + \sup_{{\bar{p}}_{X_2} \in\mathcal{P}( \textnormal{\texttt{supp}}(p_X))}  \inf_{\sigma_{B_2}} D_{\alpha}^{\mathds{M}}({\bar{\rho}}_{X_2B_2} \rVert {\bar{\rho}}_{X_2} \otimes \sigma_{B_2})\\
    &= \mathcal{L}_\alpha^{\max} (X \to B_1)_\rho + \mathcal{L}_\alpha ^{\max} (X \to B_2)_\rho,
\end{align}
where: (a) additivity holds from Lemma~\ref{lemma:Super-additivity_measure} since the conditional independency of $B_1$ and $B_2$ given $X$ in $B_1-X-B_2$ implies that the optimal measurement $\Pi = \Pi_1 \otimes \Pi_2$ of joint measured R\'enyi divergence $D_{\alpha}^{\mathds{M}}(\rho_{XB_1B_2} \rVert \rho_{X} \otimes \sigma_{B_1} \otimes \sigma_{B_2})$ is separable; (b) is from that optimal probability distributions $p_{X_1}, {p}_{X_2}$ may differ for the two corresponding subproblems.
    
\end{proof}

\section{Proof of Theorem~\ref{thm:regularizedMI}} \label{app:REG}
We start with the equivalence of regularized measured R\'enyi divergence and sandwiched R\'enyi divergence. Next, we prove the equivalence of regularized measured Arimoto information and sandwiched R\'enyi information (under the $\alpha$-tilted distribution) in Lemma~\ref{lemma:regularized_arimoto_sandwiched}. Finally, with these building blocks at hand, we can prove Theorem \ref{thm:regularizedMI} and Theorem \ref{thm:regularized_capacity}.
\begin{lemma}
    [Regularized measured R\'enyi divergence {\cite[Eq.~(4.34)]{hiai2017different}}] \label{lemma:regular_div}
    Given density matrices $\rho$ and $\sigma$. 
    For $\alpha \geq \frac{1}{2}$, we have
    \begin{equation}
        \lim_{n \rightarrow \infty} \frac{1}{n} D^{\mathds{M}}_\alpha\left( \rho^{\otimes n} \rVert \sigma^{\otimes n} \right) = D^*_{\alpha}(\rho \rVert \sigma).
    \end{equation}    
\end{lemma}
For $\alpha = 1$ case, see Ref.~\cite[Thm.~2.1]{hiai1991proper}, for $\alpha > 1$, see Ref.~\cite[Thm.~3.7]{mosonyi2014quantum}, and for $[\frac{1}{2}, 1)$, see Ref.~\cite[Prop.~8]{Hayashi2016correlation}.  


Beigi and Tomamichel \cite{beigi2023lower} proved the equivalence of regularized measured R\'enyi information and sandwiched R\'enyi information, which we list as Lemma~\ref{lemma:regularized_renyi_sandwiched}.
However, it is the measured Arimoto information that plays a role in characterizing $\alpha$-leakage $\mathcal{L}_{\alpha}(X\to B)_{\rho}$ in this paper. Lemma~\ref{lemma:regularized_arimoto_sandwiched} below shows that its regularization is given by the sandwiched R\'enyi information, albeit evaluated with the $\alpha$-tilted distribution $p_X^{(\alpha)}$ instead of the original input distribution $p_X$.

\begin{lemma}[Regularized measured R\'enyi information {\cite[Lemma~6]{beigi2023lower}}]
\label{lemma:regularized_renyi_sandwiched}
For a c-q state $\rho_{XB}$ and $\alpha \geq \frac{1}{2}$, we have
\begin{equation}
    \lim_{n\rightarrow\infty}\frac{1}{n}I^{\mathds{M}}_{\alpha}\left(X^n : B^n\right)_{\rho^{\otimes n}}=I^*_\alpha(X:B)_{\rho}.
\end{equation}
\end{lemma}

\begin{lemma}[Regularized measured Arimoto information]
\label{lemma:regularized_arimoto_sandwiched}

For a c-q state $\rho_{XB} = \sum_{x\in\mathcal{X}} p_X(x)|x\rangle \langle x|\otimes \rho_B^x$ and $\alpha \geq \frac{1}{2}$, we have
\begin{equation}\label{eq:regularized_arimoto}
    \lim_{n\rightarrow\infty}\frac{1}{n}I^{\mathrm{A},\mathds{M}}_{\alpha}\left(X^n : B^n\right)_{\rho^{\otimes n}}=I^*_\alpha(X:B)_{\rho^{(\alpha)}},
\end{equation}
where 
$\rho^{(\alpha)}_{XB} \equiv \sum_x p_X^{(\alpha)}(x)\ket{x}\!\!\bra{x} \otimes \rho_B^x$ and 
$p_X^{(\alpha)}$ is the $\alpha$-tilted distribution (Def.~\ref{def:tilted-alpha}).

\end{lemma}
\begin{proof}
    For $\alpha \geq \sfrac{1}{2}$, we can expand the definition of regularized measured Arimoto information as
\begin{align}
    &\quad \lim_{n\rightarrow \infty}\frac{1}{n}I^{\mathrm{A}, \mathds{M}}_{\alpha}\left(X^n : B^n\right)_{\rho^{\otimes n}}\\        &=\lim_{n\rightarrow\infty}\frac{1}{n}(H_{\alpha}(X^{n})_{p}-H_{\alpha}^{\mathds{M}}(X|B)_{\rho^{\otimes n}})\\
    &\overset{(\textnormal{a})}{=}H_{\alpha}(X)_{p}+\lim_{n\rightarrow\infty}\frac{1}{n}\inf_{\sigma_{B^n} \in \mathcal{S}(\mathcal{H}_{B^n} ) } D^{\mathds{M}}_{\alpha}(\rho_{XB}^{\otimes n}\rVert\mathds{1}_{X}^{\otimes n}\otimes \sigma_{B^n})\\
    &\overset{(\textnormal{b})}{=}H_{\alpha}(X)_{p}+\lim_{n\rightarrow\infty}\frac{1}{n}\inf_{\sigma_{B^n} \in \mathcal{S}(\mathcal{H}_{B}^{\otimes n } ) } D^{\mathds{M}}_{\alpha}(\rho_{XB}^{\otimes n}\rVert\mathds{1}_{X}^{\otimes n}\otimes \sigma_{B^n})\\
    &\overset{(\textnormal{c})}{=}H_{\alpha}(X)_{p}+\inf_{\sigma_B}D^*_{\alpha}(\rho_{XB}\rVert\mathds{1}_{X}\otimes \sigma_{B})\\
    &\overset{(\textnormal{d})}{=}\frac{-1}{\alpha-1}\log\sum_{x\in \mathcal{X}} p_{X}(x)^{\alpha}+\inf_{\sigma_B}\frac{1}{\alpha-1}\log\sum_{x\in \mathcal{X}}p_{X}(x)^{\alpha}{Q}^*_\alpha(\rho^{x}_{B}\rVert\sigma_{B})\label{lemma7:dir}\\
    &\overset{(\textnormal{e})}{=} \inf_{\sigma_B}\frac{1}{\alpha-1}\log\sum_{x\in \mathcal{X}}\frac{p_{X}(x)^{\alpha}}{\sum_{x\in \mathcal{X}}p_{X}(x)^{\alpha}}{Q}^*_\alpha(\rho^{x}_{B}\rVert\sigma_{B})\\
    &=\inf_{\sigma_{B}}\frac{1}{\alpha-1}\log\mathds{E}_{x\sim p_X^{(\alpha)}}\left[ {Q}^*_\alpha(\rho^{x}_{B}\rVert\sigma_{B})\right] = I^*_\alpha(X:B)_{\rho^{(\alpha)}} \label{lemma7:end}
\end{align}
where the first term of (a) follows from the additivity of $H_{\alpha}$ and $p_{X^n} = p_X^{\otimes n}$. 
The second term of (b) first follows from the observation that the minimizer of $\sigma_{B^n}$ is invariant under permutations of subsystems, which is applied in \cite[Lemma 6]{beigi2023lower}; then, by quantum de Finetti theorem in \cite{Caves2002}, the state $\sigma_B^n$ can be substituted by $\sigma_B^{\otimes n}$ with sub-linear cost, which vanishes when $n\rightarrow\infty$. (c) is from Lemma~\ref{lemma:regular_div}. In (d), for density matrices $\rho$ and $\sigma$, the sandwiched quasi R\'enyi divergence is defined as
\begin{subnumcases}
    {{Q}^*_\alpha(\rho\rVert\sigma):=}\Tr
    \left[ (\sigma^{\frac{1-\alpha}{2\alpha}}\rho\sigma^{\frac{1-\alpha}{2\alpha}})^{\alpha}\right], &
for $\alpha\in(0, 1)$ or for $\alpha\in(1,\infty)$, $\textnormal{\texttt{supp}}(\rho) \subseteq \textnormal{\texttt{supp}}(\sigma)$; \\
\infty, & otherwise.

\end{subnumcases}
(e) is from direct-sum property of sandwiched R\'enyi divergence, and one can refer to \cite[Proposition 7.29]{markwilde2021quantumcommunication}.
\end{proof}

Thus, we obtain Theorem~\ref{thm:regularizedMI} by combining Lemma~\ref{lemma:alpha-tilted}, Lemma~\ref{lemma:regularized_arimoto_sandwiched}, and Theorem~\ref{thm:alpha_leakage}.

\section{Proof of Proposition~\ref{prop:regularized_capacity}} \label{app:proof_regularized_capacity}

The first equality of \eqref{eq:regC} directly follows from Theorem~\ref{thm:max_alpha_leakage}.
    We establish the second equality of \eqref{eq:regC} as follows.
    Throughout the proof, 
    we denote by 
    \begin{align}
    \bar{\rho}_{X^n B^n}^n = \sum_{x^n \in \mathcal{X}^n} \bar{p}_{X^n} |x^n\rangle \langle x^n| \otimes \rho_{B^n}^{x^n}
    \end{align}
    for any input distribution $\bar{p}_{X^n} \in \mathcal{P}(\mathcal{X}^n)$ as a dummy variable in optimization.
    
    The upper bound of regularized R\'enyi capacity is straightforward.
    We have, for any $n\in\mathds{N}$,
    \begin{align}
        \lim_{n\rightarrow\infty}\frac{1}{n} C_{\alpha}^{\mathds{M}}\left(\mathscr{N}^{\otimes n}_{\mathcal{X} \to B} \right) 
        &= \lim_{n\rightarrow\infty}\frac{1}{n} \sup_{\bar{p}_{X^n}} \inf_{\sigma_{B^n} \in \mathcal{S}(\mathcal{H}^{\otimes n}) } D_\alpha^\mathds{M}(\bar{\rho}_{X^n B^n} \rVert \bar{\rho}_{X^n}\otimes\sigma_{B^n})\\
        &\leq 
        \lim_{n\rightarrow\infty}\frac{1}{n} \sup_{\bar{p}_{X^n}} \inf_{\sigma_{B^n} \in \mathcal{S}(\mathcal{H}^{\otimes n}) } D_\alpha^*(\bar{\rho}_{X^n B^n} \rVert \bar{\rho}_{X^n}\otimes\sigma_{B^n})\\
        &= \lim_{n\rightarrow\infty}\frac{1}{n} C_{\alpha}^*\left(\mathscr{N}_{\mathcal{X} \to B}^{\otimes n} \right) \\
        &\overset{(*)}{=} C_{\alpha}^*\left(\mathscr{N}_{\mathcal{X} \to B} \right), 
    \end{align}
    where the inequality comes from \eqref{eq:mea_leq_san}, and ($*$) comes from the additivity of sandwiched R\'enyi capacity \cite[Lemma~6]{gupta2015multiplicativity} for i.i.d.~c-q channels $\mathscr{N}_{\mathcal{X}\to B}^{\otimes n}$.
    
    For the lower bound, let $|\textrm{spec}(\sigma)|$ denote the number of mutually different eigenvalues of $\sigma$, and let $\mathscr{P}_M(\cdot) := \sum_i P_i \cdot P_i $ be the \emph{pinching map} with respect to Hermitian $M$ which has eigen-projections $\{P_i\}_{i}$.
    Then, we calculate
    \begin{align}
        &\quad \lim_{n\rightarrow\infty}\frac{1}{n} C_{\alpha}^{\mathds{M}}\left(\mathscr{N}^{\otimes n}_{\mathcal{X} \to B} \right) 
        = \lim_{n\rightarrow\infty}\frac{1}{n} \sup_{\bar{p}_{X^n} \in \mathcal{P}(\mathcal{X}^n) } \inf_{\sigma_{B^n} \in \mathcal{S}(\mathcal{H}^{\otimes n}) }  D_\alpha^\mathds{M}(\bar{\rho}_{X^n B^n} \rVert \bar{\rho}_{X^n}\otimes\sigma_{B^n})
        \\
        &\overset{(\mathrm{a})}{\geq} \lim_{n\rightarrow\infty}\frac{1}{n} \sup_{ \bar{p}_{X} \in \mathcal{P(X)} } \inf_{\sigma_{B^n} \in \mathcal{S}(\mathcal{H}^{\otimes n}) } D_\alpha^\mathds{M}( \bar{\rho}_{XB}^{\otimes n} \rVert \bar{\rho}_X^{\otimes n} \otimes \sigma_{B^n})
        \\
        &\overset{(\mathrm{b})}{=} \lim_{n\rightarrow\infty}\frac{1}{n} \sup_{ \bar{p}_{X} \in \mathcal{P(X)} } D_\alpha^\mathds{M}( \bar{\rho}_{XB}^{\otimes n} \rVert \bar{\rho}_X^{\otimes n} \otimes \sigma_{B^n}^{\star} )
        \\
        &\overset{(\mathrm{c})}{\geq} \lim_{n\rightarrow\infty}\frac{1}{n} \sup_{ \bar{p}_{X} \in \mathcal{P(X)} }  D_\alpha^* \left(\mathscr{P}_{\bar{\rho}_X^{\otimes n} \otimes  \sigma_{B^n}^{\star}}(\bar{\rho}_{XB}^{\otimes n}) \rVert \bar{\rho}_X^{\otimes n} \otimes \sigma_{B^n}^{\star} \right)\\
        &\overset{(\mathrm{d})}{\geq} \lim_{n\rightarrow\infty}\frac{1}{n} \bigg[\sup_{ \bar{p}_{X} \in \mathcal{P(X)} } D_\alpha^*(\bar{\rho}_{XB}^{\otimes n} \rVert \bar{\rho}_X^{\otimes n} \otimes \sigma_{B^n}^{\star}) - 2\log | \mathrm{spec}(\bar{\rho}_X^{\otimes n} \otimes \sigma_{B^n}^{\star})|\bigg]
        \\
        &\overset{(\mathrm{e})}{=}
        \lim_{n\rightarrow\infty}\frac{1}{n} \bigg[\sup_{ \bar{p}_{X} \in \mathcal{P(X)} } D_\alpha^*(\bar{\rho}_{XB}^{\otimes n} \rVert \bar{\rho}_X^{\otimes n} \otimes \sigma_{B^n}^{\star}) - \log \mathrm{poly}(n) \bigg]
        \\
        &\overset{(\mathrm{f})}{\geq} \lim_{n\rightarrow\infty}\frac{1}{n} \bigg[ \sup_{ \bar{p}_{X} \in \mathcal{P(X)} } I^*_\alpha(X^n;B^n)_{\bar{\rho}^{\otimes n}} - \log \mathrm{poly}(n)\bigg]\\
        &\overset{(\mathrm{g})}{=} \lim_{n\rightarrow\infty} \sup_{\bar{p}_{X} \in \mathcal{P(X)} } I^*_\alpha(X:B)_{\bar{\rho}} - \frac{1}{n} \log \mathrm{poly}(n)\\
        &= C_{\alpha}^*\left(\mathscr{N}_{\mathcal{X} \to B} \right),
    \end{align}
    where (a) is from restricting the optimization $\bar{p}_{X^n} \in \mathcal{P}(\mathcal{X}^n)$ to i.i.d.~inputs $\bar{p}_X^{\otimes n}$;
    in (b) we denote $\sigma_{B^n}^{\star} \in \mathcal{S}(\mathcal{H}^{\otimes n} )$ as the permutation-invariant minimizer to the map $\sigma_{B^n} \mapsto D_\alpha^\mathds{M}(\bar{\rho}_{XB}^{\otimes n} \rVert \bar{\rho}_X^{\otimes n}\otimes\sigma_{B^n})$
    (which depends on $\bar{p}_X$; see e.g.~\cite[Eq.~(58)]{beigi2023lower});
    (c) from data-processing inequality of applying the     
    pinching map  $\mathscr{P}_{\bar{\rho}_X^{\otimes n} \otimes \sigma_{B^n}^{\star}}$ 
    to force the resulting states on system $B^n$ to commute (see e.g.~\cite[Lemma~6]{beigi2023lower}), i.e.
    \begin{align}
    D_\alpha^\mathds{M}( \bar{\rho}_{XB}^{\otimes n} \rVert \bar{\rho}_X^{\otimes n} \otimes \sigma_{B^n}^{\star} )
    \geq D_\alpha^{\mathds{M}}\left(\mathscr{P}_{\bar{\rho}_X^{\otimes n} \otimes \sigma_{B^n}^{\star}}(\bar{\rho}_{XB}^{\otimes n}) \rVert \bar{\rho}_X^{\otimes n} \otimes \omega_{B^n} \right)
    = D_\alpha^{*}\left(\mathscr{P}_{\bar{\rho}_X^{\otimes n} \otimes \sigma_{B^n}^{\star}}(\bar{\rho}_{XB}^{\otimes n}) \rVert \bar{\rho}_X^{\otimes n} \otimes \sigma_{B^n}^{\star} \right)\!.
    \end{align}
    
    (d) is because measured R\'enyi divergence can be lower-bounded by sandwiched R\'enyi divergence subtracted by a logarithmic number of spectrum \cite[Lemma~3]{Hayashi2016correlation}:
    \begin{subnumcases}
        {D^\mathds{M}_\alpha(\rho\rVert\sigma) \geq D^*_\alpha(\mathscr{P}_{\sigma}(\rho) \rVert \sigma) \geq D^*_\alpha(\rho\rVert\sigma) - } \log |\mathrm{spec}(\sigma)|, & if $\alpha \in [0, 2]$;\\
        2 \log |\mathrm{spec}(\sigma)|, & if $\alpha > 2$,
    \end{subnumcases}
    which results from pinching inequality \cite[Lemma~3.10] {hayashi2016quantum}:
    \begin{equation}
        \rho \leq |\mathrm{spec}(\sigma)| \mathscr{P}_\sigma(\rho);
    \end{equation}
    
    (e) is from the number of distinct eigenvalues of the permutation-invariant minimizer\\ $|\mathrm{spec}(\bar{\rho}_X^{\otimes n} \otimes \sigma_{B^n}^{\star})|$ grows polynomially with $n$, since from Schur duality, denoting $d$ as the dimension of finite-dimension Hilbert space $\mathcal{H}_X \otimes \mathcal{H}_B$, and let $\Lambda_n =\{ \lambda=(\lambda_{1},...,\lambda_{d})|\lambda_{1}\geq...\geq\lambda_{d}\geq 0 \text{ and }\sum_{i=1}^n \lambda_i=n \}$ be partitions of $n$ into $\leq d$ parts, $\bar{\rho}_X^{\otimes n} \otimes \sigma_{B^n}^{\star}$ can be represented in Schur basis $P_\lambda$ and $Q^d_\lambda$ (see e.g.~\cite[Eq.~(5.16), (5.18)]{harrow2005applications}\cite[Eq.~(48)]{berta2021composite}):
    \begin{equation}
        \sum_{\lambda \in \Lambda_n} \ket{\lambda}\!\!\bra{\lambda} \otimes \sigma_{Q_\lambda^d} \otimes \mathds{1}_{P_\lambda},
    \end{equation}
    where one can upper-bound these quantities as follows \cite[Sec.~6.2]{harrow2005applications}:
    \begin{align}
        |\Lambda_n| &\leq (n + 1)^d \in \mathrm{poly}(n);\\
        \mathrm{dim~} Q_\lambda^d &\leq (n + d)^{\frac{d(d - 1)}{2}} \in \mathrm{poly}(n).
    \end{align}
    Note that this holds for any $\bar{p}_X \in \mathcal{P}(\mathcal{X})$.
    We remark that a similar analysis of this step was also used in \cite[Lemma~2.4]{berta2021composite}.
    
    (f) follows from the minimization in the definition of measured R\'enyi information; 
    (g) from additivity of sandwiched R\'enyi information under product states \cite[Lemma~4.8]{mosonyi2017strong}. 
    We remark that a similar technique was applied in the proof of \cite[Thm.~5.14]{mosonyi2017strong}.

    Since both the upper bound and the lower bound are given by $C_{\alpha}^*\left(\mathscr{N}_{\mathcal{X} \to B} \right)$, this concludes the proof.

\end{appendices}

\clearpage


{
\bibliographystyle{myIEEEtran}
\bibliography{refs.bib}
}

\end{document}